\documentclass[12pt]{article}

\usepackage{epsfig} 
\usepackage{amsmath,amssymb} 
\newcount\minutes 
\newcount\scratch 
 
\def\timestamp{%
\scratch=\time 
\divide\scratch by 60 
\edef\hours{\the\scratch} 
\multiply\scratch by 60 
\minutes=\time 
\advance\minutes by -\scratch 
---$\,$\hours:\null 
\ifnum\minutes< 10 0\fi 
\the\minutes} 

\newlength{\fdagwidth}
\newlength{\diagupwidth}
\newlength{\stepback}
\newcommand{\fdag}[2][\diagup]{\text{$#2$\settowidth{\fdagwidth}{$#2$}\settowidth{\diagupwidth}{$#1$}\setlength{\stepback}{0.5\fdagwidth}\hspace{-\stepback}\hspace{-0.5\diagupwidth}$#1$\hspace{\stepback}\hspace{-0.5\diagupwidth}}}

\def\lsim{\mathrel{\raisebox{-.6ex}{$\stackrel{\textstyle<}{\sim}$}}}
\def\gsim{\mathrel{\raisebox{-.6ex}{$\stackrel{\textstyle>}{\sim}$}}}
\def\slashiv#1{#1\llap{\sl/}}

\begin{document}
\begin{titlepage} 
\nopagebreak  
{\flushright{ 
        \begin{minipage}{5cm}
         KA--TP--04--2007  \\       
         SFB/CPP--07--10  \\
        {\tt hep-ph/0703202}\hfill \\ 
        \end{minipage}        } 
 
} 
\vfill 
\begin{center} 
{\LARGE \bf 
 \baselineskip 0.9cm 
 Higgs plus two jet production via gluon fusion\\
 as a signal at the CERN LHC
} 
\vskip 0.5cm  
{\large   
G.~Kl\"amke and D.~Zeppenfeld
}   
\vskip .2cm  
{{\it Institut f\"ur Theoretische Physik, 
  Universit\"at Karlsruhe, P.O.Box 6980, 76128 Karlsruhe, Germany}
} 
 
 \vskip 
1.3cm     
\end{center} 
 
\nopagebreak 
\begin{abstract}
Higgs boson production in association with two tagging jets will be
mediated by electroweak vector boson fusion and by gluon fusion
processes at the CERN LHC. The characteristic distributions for the
gluon fusion process are analyzed for the $H\to W^+W^-$ signal at 
$m_H=160$~GeV, with subsequent leptonic decay of the $W$-pair. The
dominant backgrounds from top-quark pair production, $WWjj$ production
and vector boson fusion processes can be suppressed to a level of
$S/B\approx 1/4$, yielding a highly significant gluon fusion signal with
30~fb$^{-1}$.
Analysis of the azimuthal angle correlations of the two jets provides
for a direct measurement of the CP-nature of the $Htt$ Yukawa coupling
which is responsible for the effective $Hgg$ vertex.
\end{abstract} 
\vfill 
\vfill 
\end{titlepage} 
\newpage               
%
%

\section{Introduction}

Higgs boson production in association with two jets has emerged as a 
promising channel for Higgs boson discovery~\cite{ATLAS,ATLAS_VBF,CMS,VBF:H} 
and for the study of Higgs boson properties~\cite{VBF:C,VBF:CP} at 
the CERN Large Hadron Collider (LHC). Interest has concentrated on
vector-boson-fusion (VBF), i.e. the weak process $qq\to qqH$ which is 
mediated by $t$ channel exchange of a $W$ or $Z$, with the Higgs boson 
being radiated off this weak boson. The VBF production cross section measures
the strength of the $WWH$ and $ZZH$ couplings, which, at tree
level, require a vacuum expectation value for the scalar field. Hence the 
VBF channel is a sensitive probe of the Higgs mechanism as the source
of electroweak symmetry breaking.

\begin{figure}[htb]
\begin{center}
\begin{tabular}{ccc}
\includegraphics[scale=0.68]{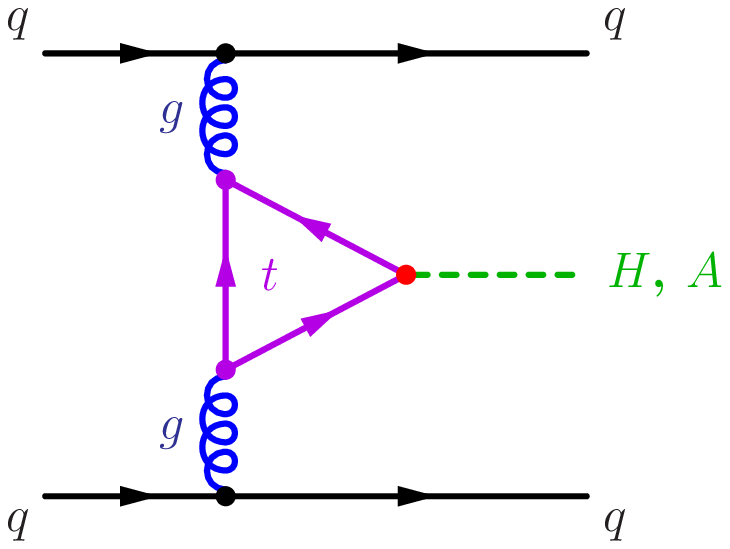}&
\includegraphics[scale=0.68]{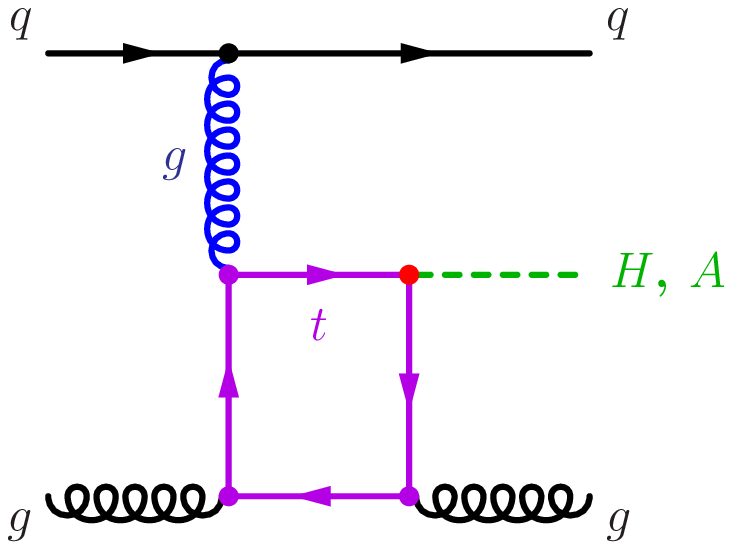}&
\includegraphics[scale=0.72]{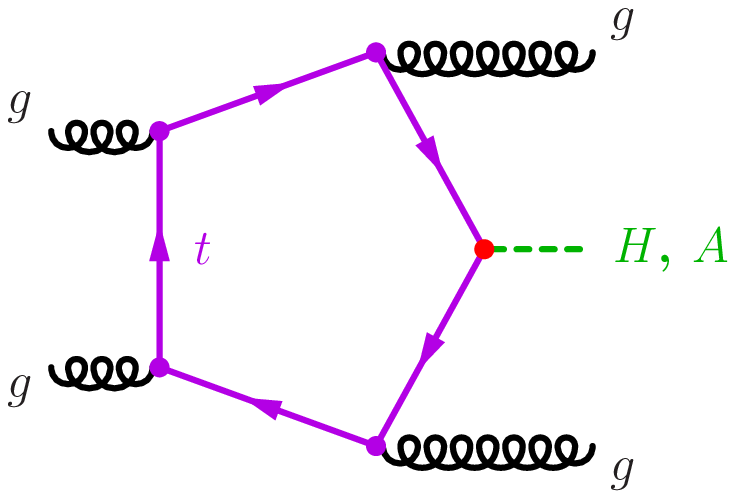}\\
\end{tabular}
\end{center}
\caption[]{\label{fig:tmunu} Feynman graphs contributing to $pp\to Hjj$. }
\end{figure}

Another prominent source of $Hjj$ events are second order real emission 
corrections to the gluon fusion process. Such corrections were first 
considered in Ref.~\cite{Kauffman:1996ix, Kauffman:1998yg} in the large 
top mass limit and have subsequently been evaluated for arbitrary quark masses
in the loops which induce the effective coupling of the Higgs
boson to gluons~\cite{DelDuca:2001eu}. Some representative Feynman
graphs are shown in Fig.~\ref{fig:tmunu}. In the $m_t\to\infty$ limit, also
the NLO QCD corrections have recently been
calculated~\cite{Campbell:2006xx}. 

For a SM Higgs boson, the generic $Hjj$ cross section from gluon fusion
can somewhat exceed the VBF cross section of a few
pb~\cite{DelDuca:2001eu}. This raises the question whether 
gluon fusion induced $Hjj$ events can be used as a source of information
for measuring Higgs boson properties. For the VBF process, the most
promising Higgs signal arises for Higgs boson masses around $W$-pair
threshold in the channel $pp\to HjjX,\; H\to W^+W^-\to
l^+l^-\slashiv{p}_T$~\cite{ATLAS_VBF,VBF:H}. This prompts us
to investigate the gluon fusion contribution to this channel for a Higgs
boson mass of $m_H=160$~GeV. We analyze potential background processes,
in particular the production of top-quark pairs in association with
additional jets, QCD induced $WWjj$ events and VBF processes, and we
show in a parton level analysis 
that the gluon fusion induced Higgs production can be isolated as a
highly significant signal with 30~fb$^{-1}$ of LHC data and with a
signal to background ratio of about one to four. 

The resulting signal is large enough to derive Higgs boson properties
from distributions. In this paper we focus on the CP properties of the 
Yukawa couplings to fermions, which are given by
\begin{equation}
{\cal L}_Y = y_f H\bar \psi_f \psi_f +i \tilde y_f A\bar
\psi_f\gamma_5\psi_f,
\label{eq:Yuk}
\end{equation}
where $H$ and $A$ denote (pseudo)scalar Higgs fields which couple
to fermions $f= t, b, \tau$ etc. Via these Yukawa couplings, quark
loops induce effective couplings of the Higgs boson to gluons. In our
numerical analysis we consider couplings of SM strength, $y_f=\tilde
y_f = m_f/v = y^{SM}$. In this case the quark loops are dominated by
the top quark, and the Higgs gluon coupling can be described by the
effective Lagrangian~\cite{Kauffman:1996ix, Kauffman:1998yg}
\begin{equation}
{\cal L}_{\rm eff} = 
\frac{y_t}{y_t^{SM}}\cdot\frac{\alpha_s}{12\pi v} \cdot H \,G_{\mu\nu}^a\,G^{a\,\mu\nu} +
\frac{\tilde y_t}{y_t^{SM}}\cdot\frac{\alpha_s}{16\pi v} \cdot A \,
G^{a}_{\mu\nu}\,G^{a}_{\rho\sigma}\varepsilon^{\mu\nu\rho\sigma}\;,
\label{eq:ggS}
\end{equation}
where $G^{a}_{\mu\nu}$ denotes the gluon field strength. The effective
Lagrangian approximation provides an excellent description of the full
results for $Hjj$ production, provided one considers modest jet
transverse momenta, $p_{Tj} \lsim m_t$, and Higgs boson masses well
below the top quark pair production threshold~\cite{DelDuca:2001eu}.
We employ the effective Lagrangian description throughout this paper. 
From the effective Lagrangian emerge $Hgg$, $Hggg$ and also $Hgggg$ 
vertices, which correspond to triangle, box and pentagon top quark 
loops as in Fig.~\ref{fig:tmunu}.

The structure of the left diagram in Fig.~\ref{fig:tmunu} is very
similar to the process of Higgs production in vector boson 
fusion~\cite{VBF:H}. In Ref.~\cite{Hankele:2006ma} it
was shown that the distribution of the azimuthal angle between the two
jets in $Hjj$ events can be used to determine the tensor structure of
the $HVV$ coupling ($V=W^\pm\,,Z$). The same method can be applied to 
Higgs$ + 2$~jet production in gluon fusion. Here, the azimuthal angle 
distribution is sensitive to the tensor structure of the effective 
$Hgg$ coupling, which is determined by the CP-structure of the top Yukawa
coupling. More precisely, neglecting terms which vanish upon contraction
with the conserved quark currents, the tensor structure of the $Hgg$ vertex
which emerges from Eq.(\ref{eq:ggS}) is given by
\begin{equation}
T^{\mu\nu} = a_2\,(q_1\cdot q_2\, g^{\mu\nu} - 
q_1^\nu q_2^\mu) + a_3\, \varepsilon^{\mu\nu\rho\sigma}q_{1\rho}q_{2\sigma}\,,
\label{eq:Tmunu}
\end{equation}
where $q_1$ and $q_2$ are the four-momenta of the two gluons. The scalar form factors
$a_2$, $a_3$ are directly related to the Yukawa interactions of Eq.~(\ref{eq:Yuk}),
\begin{equation}
a_2 = \frac{y_t}{y_t^{SM}}\cdot\frac{\alpha_s}{3\pi v}\,,\qquad
a_3 = -\frac{\tilde y_t}{y_t^{SM}}\cdot\frac{\alpha_s}{2\pi v}
\label{eq:a2a3}
\end{equation}
Note that $|a_3|=\frac{3}{2}\cdot|a_2|$ for $y_t=\tilde y_t$. Therefore the
cross section for the case of a purely CP-odd Yukawa interaction will
be about $1.5^2=2.25$ times larger than the Standard Model cross section.
In contrast to vector boson fusion, there are
additional contributions from Higgs plus three and four gluon
couplings. These may dilute the sensitivity to the structure of the $Hgg$
vertex and we need to find the regions of phase space with the best
analyzing power for differences between CP-even and CP-odd Yukawa
couplings.

The paper is organized as follows. In Section~\ref{sec:tools} we briefly
describe the parton level Monte Carlo programs with which we determine 
the relevant cross sections. All calculations are done at tree level.
We then determine the characteristic distributions of the gluon fusion
signal and of the various backgrounds in Section~\ref{sec:xLHC} and
devise cuts which provide a reasonable signal to background ratio and a
high statistical signal significance. The measurement of the CP
structure of the top Yukawa coupling is addressed in
Section~\ref{sec:azimjj}. We determine the analyzing power of the
azimuthal angle correlations between the tagging jets and find that it
becomes more pronounced for larger rapidity separations of the jets.
Final conclusions are drawn in Section~\ref{sec:conclusions}.

\section{Calculational tools}
\label{sec:tools}

We consider Higgs$+2$~jet production in gluon fusion (GF) with the Higgs
decaying into a pair of $W$ bosons, which, in turn decay leptonically
into electrons and muons ($\ell^\pm = e^\pm, \mu^\pm$) and the
associated neutrinos: $pp\to H jjX,\; H \to W^+W^-  \to \ell^+ \ell^- \nu
\bar{\nu}$.
The dominant backgrounds are from top-pair production,
$pp\to t\bar{t}X$, and from $t\bar{t}$ production with additional 
jets, $pp\to t\bar{t} jX$, $pp\to t\bar{t} jjX$. 
Another background is $W$ pair production with two accompanying jets, 
$pp\to W^+W^- jjX$. This can be a QCD induced process of order
$\alpha^2\alpha_s^2$, or it may arise from vector boson fusion,
i.e. electroweak processes of the type $qq\to qqW^+W^-$ at order
$\alpha^4$. We do not
consider backgrounds from $Zjj$, $Z\to \tau^+\tau^-$ and 
from $b\bar{b}jj$ production because they have been shown to be small in
the analysis of Ref.~\cite{VBF:H}. 
All signal and background cross sections are determined in terms of full
tree level matrix elements for the contributing subprocesses and are
discussed in more detail below. Detector resolution effects are
neglected in the following.

For all our numerical results we simulate $pp$ collisions at a center of
mass energy of $\sqrt{s} = 14$ TeV.  
Standard Model parameters are set to $\sin^2{\theta_W} = 0.23105$, $M_Z
= 91.187$~GeV and $G_F = 1.16637\cdot 10^{-5}\,{\rm GeV}^{-2}$, which
translates into $M_W = 79.962$~GeV and $\alpha(M_Z) 
= 1/128.92$ when using the tree-level relations between these input
parameters.  For all QCD effects, the running of the strong coupling
constant is evaluated at leading order, with $\alpha_s(M_Z) =0.1298$. We
employ CTEQ6L1 parton distribution functions throughout~\cite{cteq6}.

\subsection{The $H + jj$ signal process}
The production of a Higgs boson in gluon fusion in association with two jets, at order
$\alpha_s^4$, can proceed via the subprocesses~\cite{Kauffman:1996ix, Kauffman:1998yg}
\begin{equation}
qq'\rightarrow qq'H\,,\qquad qg\rightarrow qgH\,,\qquad gg\rightarrow ggH,
\end{equation}
and all crossing related processes. 
The calculation of this process is based on the work of Ref.~\cite{DelDuca:2001eu}.
Instead of full top quark loops we express the matrix elements in
terms of effective Higgs gluon vertices as given by
Eqs.~(\ref{eq:ggS},\ref{eq:Tmunu}). We include CP-even and CP-odd
Higgs couplings and any interference between them.
In addition the program was extended with the matrix elements for the
Higgs decay $H\to W^+W^- \to \ell^+\ell^-\nu\bar{\nu}$. 

Throughout the analysis we
use the Standard Model coupling and branching ratio for the Higgs decay into $W^+W^-$, 
$B(H\to W^+W^-)=0.912$ for a Higgs boson mass of $m_H = 160\,{\rm GeV}$. 
For the signal process, the factorization scale is chosen as
$\mu_f=\sqrt{p_{T_1}\cdot p_{T_2}}$, where $p_{T_{1/2}}$ denote the transverse
momenta of the two jets. The strong coupling constant is taken as
$\alpha_s^4 = \alpha_s^2(m_H)\cdot \alpha_s(p_{T_1})\cdot \alpha_s(p_{T_2})$.

\subsection{The QCD $t\bar{t} + jets$ backgrounds}
Given the Higgs decay signature, the main physics background to the
$\ell^\pm\ell^\mp jj \fdag{p}_T$ signal arises from $t\bar{t} + jets$
production, due to the large production cross section at the LHC and
because the branching ratio $B(t\rightarrow Wb)$ is essentially 100\%.
The basic process we consider is $pp\rightarrow t\bar{t}$, which can be
either $gg-$ or $q\bar{q}$-initiated, with the former strongly
dominating at the LHC. Real emission QCD corrections lead to $t\bar{t}+j$ and
$t\bar{t}+jj$ events. Relevant subprocesses for the latter are
\begin{equation}
q\bar{q}\rightarrow t\bar{t}g\,,\qquad 
      gg\rightarrow t\bar{t}g\,,\qquad 
q\bar{q}\rightarrow t\bar{t}q\bar{q}\,,\qquad
      gg\rightarrow t\bar{t}q\bar{q}\,,\qquad
      gg\rightarrow t\bar{t}gg
\end{equation}
and all crossing-related processes~\cite{Kauer:2002sn}.

We calculate the $t\bar{t}$, $t\bar{t}j$ and $t\bar{t}jj$ processes at leading
order. Hence, for $t\bar{t}$+jets one has to avoid the
phase space regions where the massless jets get soft or collinear with
the $b$ quarks, in order to have a finite cross section. In addition,
double counting has to be avoided when combining the three background
processes. This is achieved in the following way. For the $t\bar{t}jj$
case, we require both tagging jets to arise from massless partons in our
simulation. Similarly for $t\bar{t}j$ production, exactly one tagging jet
is allowed to arise from a $b$ quark. Finally, the $t\bar{t}$
cross section corresponds to both tagging jets arising from $b$
quarks. With this prescription there is no double counting and the
cross sections are finite when applying the cuts described below in 
Section~\ref{sec:xLHC}.

The calculation has been performed using the program of Ref.~\cite{Kauer:2002sn}.
The decays of the top quarks and $W$'s are included in the matrix
elements, taking into account all off-shell effects for the leptonic
final state~\cite{Kauer:2001sp}.
In all cases, the factorization scale is chosen as $\mu_f=\min(E_T)$ 
of the top quarks and additional jets. The overall strong coupling
constant factors for the LO $t\bar t+n$~jet cross section are
calculated as $(\alpha_s)^{n+2}=\prod^{n+2}_{i=1}\alpha_s(E_{T_i})$, where the
product runs over $n$ massless partons and the two top quarks.

\subsection{The EW $WW+jj$ background}
This background arises from $W^+W^-$ bremsstrahlung in
quark-(anti)quark scattering via $t$-channel electroweak boson exchange,
with subsequent decay $W^+W^-\rightarrow \ell^+\ell^-\fdag{p}_T$, i,e,
\begin{equation}
qq'\rightarrow qq'W^+W^-\rightarrow \ell^+\ell^- jj  \fdag{p}_T
\end{equation}
and crossing related processes. 
The process was calculated with the program 
VBFNLO~\cite{Hankele:2006ma,Figy:2003nv, Oleari:2003tc, Jager:2006cp}, 
which also allows to calculate NLO corrections to distributions. We only 
use the tree level option, however. In Ref.~\cite{Jager:2006cp} it was 
shown that NLO effects are minimized by the factorization scale choice 
$\mu=Q$ at tree level, where $Q$ is the momentum transfer of the 
$t$-channel electroweak boson. With this choice higher order QCD effects
are well below 10\%. 
This EW $WW+jj$ background also includes Higgs production in VBF. In fact,
for $m_H=160$~GeV it is dominated by the Higgs contribution, which we here
consider as a background to the observation of $Hjj$ production in gluon 
fusion. In the following we collectively refer to the VBF Higgs 
contribution as well as to continuum $WW$ production in VBF as the 
``EW $WWjj$'' background.

\subsection{The QCD $WW+jj$ background}
Real-emission QCD corrections to $W^+W^-$ production give rise to $W^+W^-jj$
events. These background processes include~\cite{Barger:1989yd}
\begin{equation}
qq'\rightarrow qq'W^+W^-\,,\qquad qg\rightarrow qgW^+W^-,
\end{equation}
which are dominated by t-channel gluon exchange, and all crossing
related processes. The calculation has been done in the framework of the
VBFNLO program by employing matrix elements that have been generated 
with MadGraph~\cite{Maltoni:2002qb}.
We call these processes collectively the ``QCD $WWjj$'' background.
The factorization scale is chosen as $\mu=\min(p_{T_1},p_{T_2})$ of the 
two final state partons. The
strong coupling constant factor is taken as $\alpha_s^2 =
\alpha_s(p_{T_1})\cdot \alpha_s(p_{T_2})$, i.e. the transverse momentum
of each additional parton is taken as the relevant scale for its production.

\section{Cross sections at the LHC}
\label{sec:xLHC}
\begin{figure}[htb]
\centerline{
\includegraphics[scale=0.41]{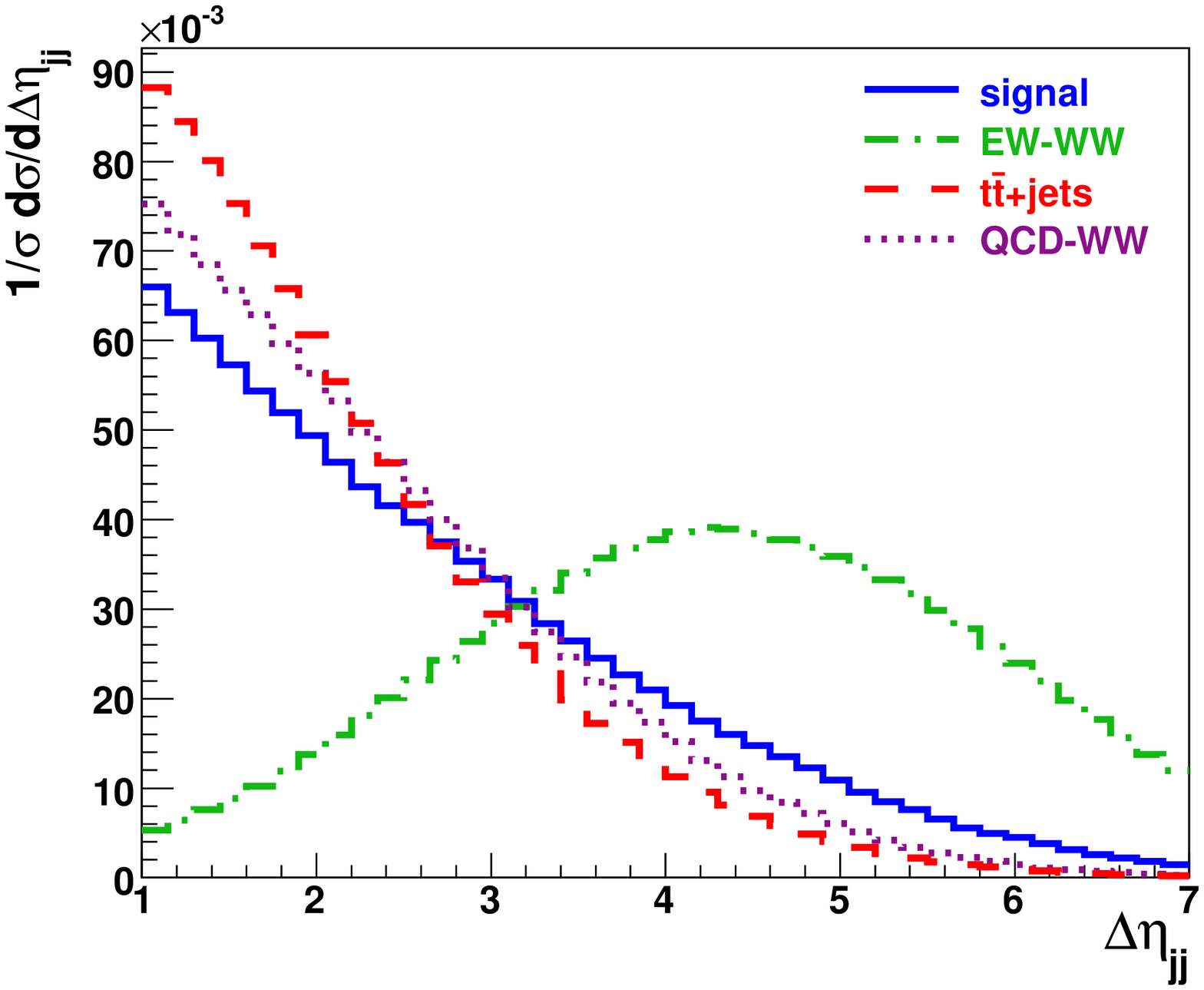}
\includegraphics[scale=0.41]{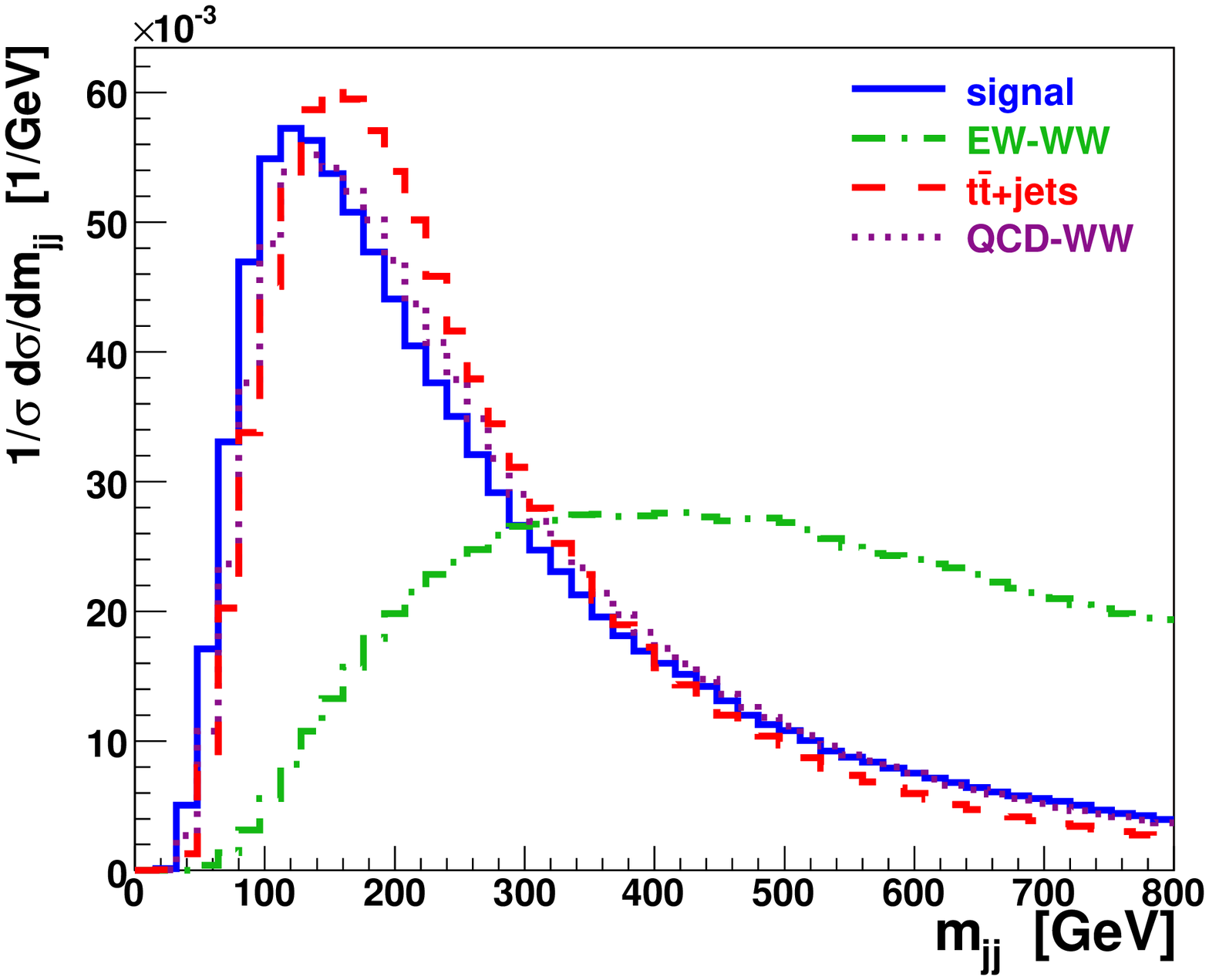}
}
\caption[]{\label{fig:etajj}  Normalized rapidity separation {\it (left)} and dijet
  invariant mass {\it (right)} distribution of the two tagging jets for
  the signal (solid) and backgrounds: EW $W^+W^-jj$ (dash-dotted),
  $t\bar{t}+jets$ (dashed) and QCD $W^+W^-jj$ (dotted). The cuts of Eq.~(\ref{eqn:incl}) are imposed.}
\end{figure}

The gluon fusion induced 
$pp\rightarrow HjjX$, $H\rightarrow W^{(*)}W^{(*)}\rightarrow
\ell^\pm\ell^\mp\nu\bar{\nu}$ signal is characterized by two high $p_T$
tagging jets and the $W$ decay leptons ($e$, $\mu$). Here, the tagging
jets are defined as the two highest momentum jets in an event. Thus, the
signal characteristic is similar to the VBF Higgs signal~\cite{VBF:H}. 
However, one cannot simply follow the same search strategy as for VBF. 
The reason is illustrated in
Fig.~\ref{fig:etajj} which shows the rapidity separation,
$\Delta\eta_{jj}=|\eta_{j_1}-\eta_{j_2}|$, of the two tagging jets
and the dijet invariant mass, $m_{jj}$, for the signal and the background
processes. The three $t\bar{t}+jets$ backgrounds have been combined for
clarity, even though their individual distributions are slightly
different. The shape of the distributions for $H + jj$ in
VBF is very different from that for $H + jj$ in GF which resembles the
background distributions. This is because the GF signal shares the
characteristics of QCD processes, which are dominated by external gluons.
So while these observables are very
powerful in the VBF analysis, they are almost useless for separating the
GF signal from the background. Therefore, we have to expect a significantly
worse signal to background ratio compared to the VBF analyses.  
In the following we optimize cuts for a Higgs boson mass of $m_H=160$ GeV.
For the calculation of the signal and background processes we impose the following minimal cuts: 
\begin{equation*}
p_{Tj} > 30\, {\rm GeV},\qquad|\eta_j| < 4.5,\qquad |\eta_{j_1}-\eta_{j_2}| > 1.0
\end{equation*}
\begin{equation}
p_{T\ell} > 10\, {\rm GeV},\qquad|\eta_\ell| < 2.5,\qquad \Delta R_{j\ell} = \sqrt{(\eta_j-\eta_\ell)^2+(\Phi_j-\Phi_\ell)^2} > 0.7
\label{eqn:incl}
\end{equation}
Thus, the jets are required to have a transverse momentum of more than 30
GeV and a rapidity below 4.5 to be detected in the hadronic
calorimeter. In the same way the two charged leptons are required to
have at least 10 GeV of transverse momentum and they should be 
sufficiently central in order to provide tracking information. 
Furthermore the jets and leptons are forced to be well separated 
from each other. Below we will argue for a substantially higher $p_T$ 
threshold for the charged leptons. Hence, standard LHC lepton triggers
will have a have a very high efficiency for these $W$-pair events.

At the level of the inclusive cuts of Eq.~(\ref{eqn:incl}) we require only a
modest rapidity separation of $\Delta\eta_{jj} > 1$ for the two tagging
jets which are defined as the two highest $p_T$ jets of an event. In 
contrast to the VBF studies, we do not require the two leptons to lie between 
these two tagging jets. Instead we will focus on angular and mass cuts of
the Higgs decay products to isolate the signal.

The cross sections resulting for the cuts of Eq.~(\ref{eqn:incl}) 
are shown in the first line of
Table~\ref{tab:xs}. The signal cross section of 115 fb (which includes
the branching ratios into leptons) is quite sizeable. The QCD $WWjj$
cross section is about 3 times higher whereas the VBF process reaches
2/3 of the signal rate. The worst source of background arises from the
$t\bar{t}$ processes, however, with a combined cross section of about 18 pb. 

\begin{figure}[htb]
\centerline{
\includegraphics[scale=0.41]{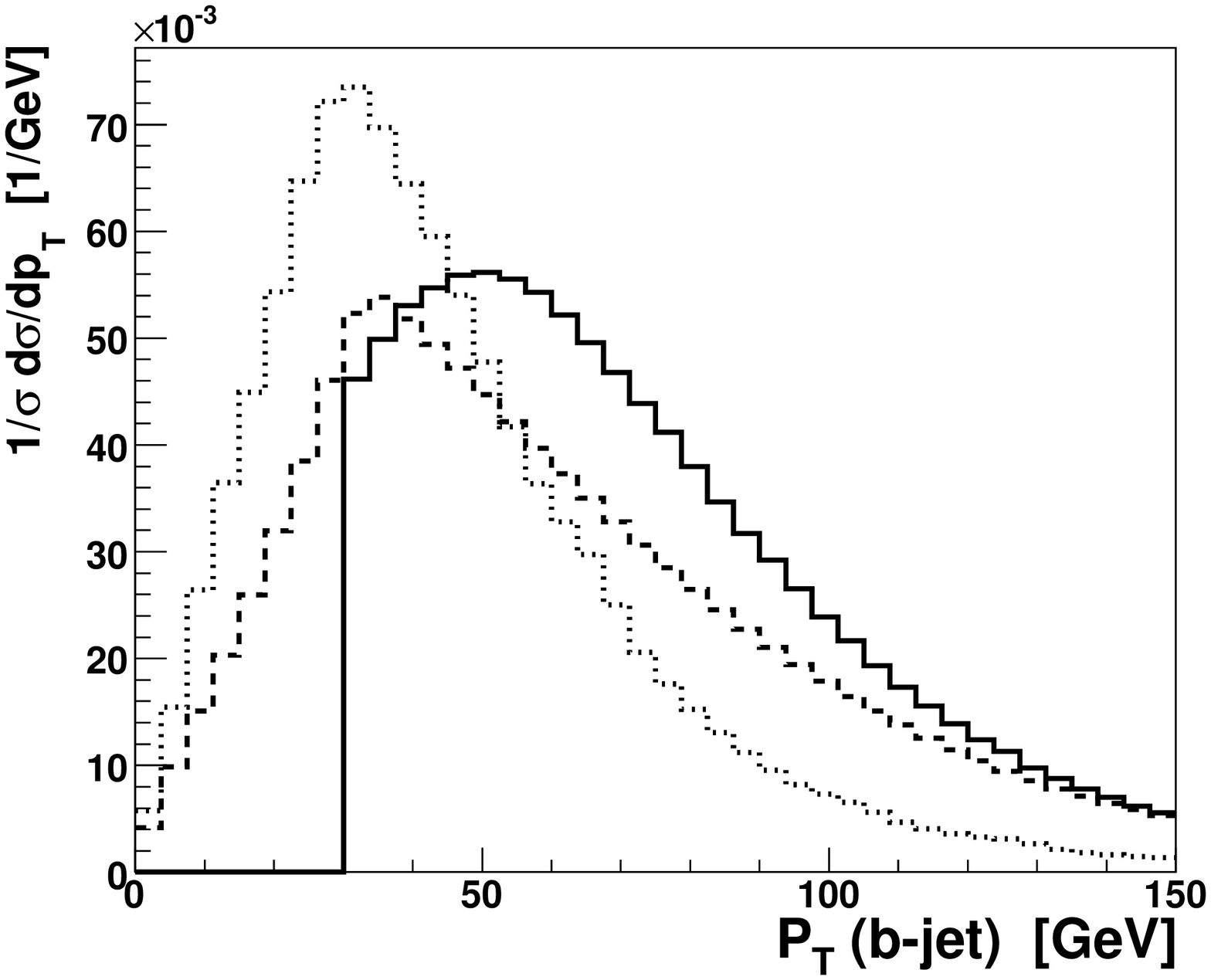}
\includegraphics[scale=0.41]{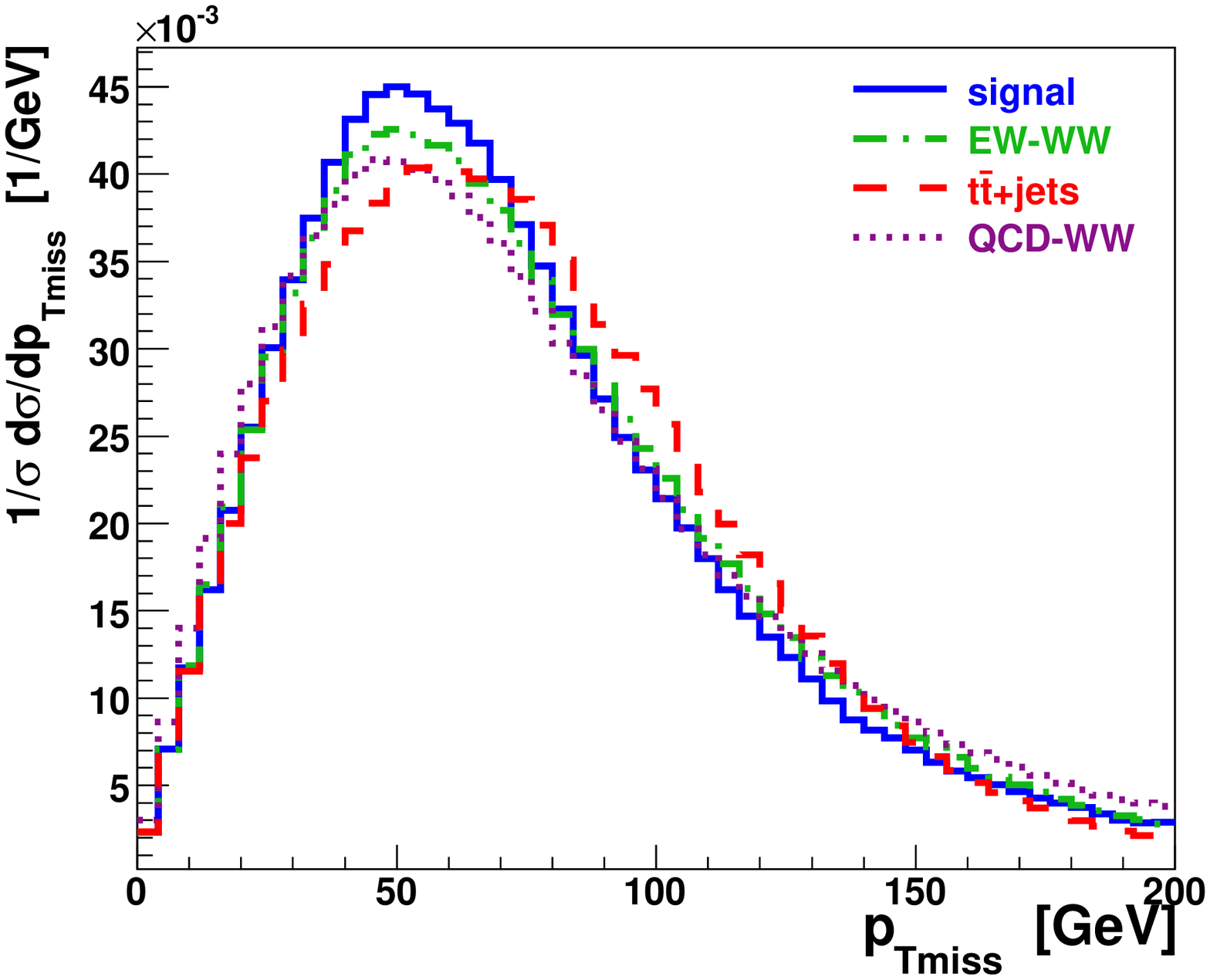}
}
\caption[]{\label{fig:pt} {\it Left:} Normalized transverse momentum distribution
  of the $b$-quarks for $t\bar{t}$ (solid), $t\bar{t}+j$ (dashed) and $t\bar{t}+jj$
  (dotted). {\it Right:} Normalized distribution of the missing
  transverse momentum for signal and backgrounds as in
  Fig.~\ref{fig:etajj}. The cuts of Eq.~(\ref{eqn:incl}) are imposed.} 
\end{figure}

In order to reduce this large $t\bar{t}$ background it is necessary to
make use of a $b$-veto, that is to discard all events where one or both jets
are tagged as $b$ jets. We allow for an overall mistagging probability 
of 10\% for light partons with $p_T > 30$ GeV and $|\eta| < 2.4$, which 
leads to an acceptable reduction of less than 20\% for the signal and 
all non-$t\bar{t}$ backgrounds. 
We use the results of the CMS analysis of Ref.~\cite{Weiser:2006md} 
for our assumptions on
$b$-veto efficiencies and mistagging probabilities. For a 10\% mistagging 
probability per jet one finds $b$-veto efficiencies in the range of 
60\% - 75\%, depending on jet transverse momentum and rapidity
$(p_T,\eta)$ as shown in Table~\ref{tab:bveto}. 
\begin{table}[htb]
  \caption{The assumed $b$-tagging efficiencies as a function of jet
    $p_T$ and rapidity. From Ref.~\cite{Weiser:2006md}}
\begin{center}
\begin{tabular}{|c|c|c|}
 \cline{2-3}
\multicolumn{1}{c|}{}  & $30\,{\rm GeV} < p_T < 50\,{\rm GeV}$ & $p_T > 50\,{\rm GeV}$  \\
\hline
$1.4 < |\eta| < 2.4$ & 60\% & 70\% \\
$|\eta| < 1.4$ & 65\% & 75\%\\
\hline
\end{tabular}
\end{center}
\label{tab:bveto}
\end{table}
With the $b$-veto, the top backgrounds are reduced by factors of 3 to 8 
as shown
in line 2 of Table~\ref{tab:xs}. The $b$-veto is less efficient for the
$t\bar{t}jj$ and $t\bar{t}j$ processes than for the $t\bar{t}$ process
since, by definition of the tagging jets, the $p_T$ of the $b$ quark 
becomes smaller the more massless jets are radiated. This is illustrated 
in the left plot of Fig.~\ref{fig:pt}.
The curve for the $t\bar{t}$ process shows the cut at 30 GeV of
Eq.~\ref{eqn:incl} because in this case both b-jets are tagging
jets. For the $t\bar{t}j$ and $t\bar{t}jj$ processes either one or 
both $b$-jets are distinct from the two tagging jets
and therefore the $p_{T}$ distributions continue below 30~GeV.
 
\begin{figure}[htb]
\centerline{
\includegraphics[scale=0.41]{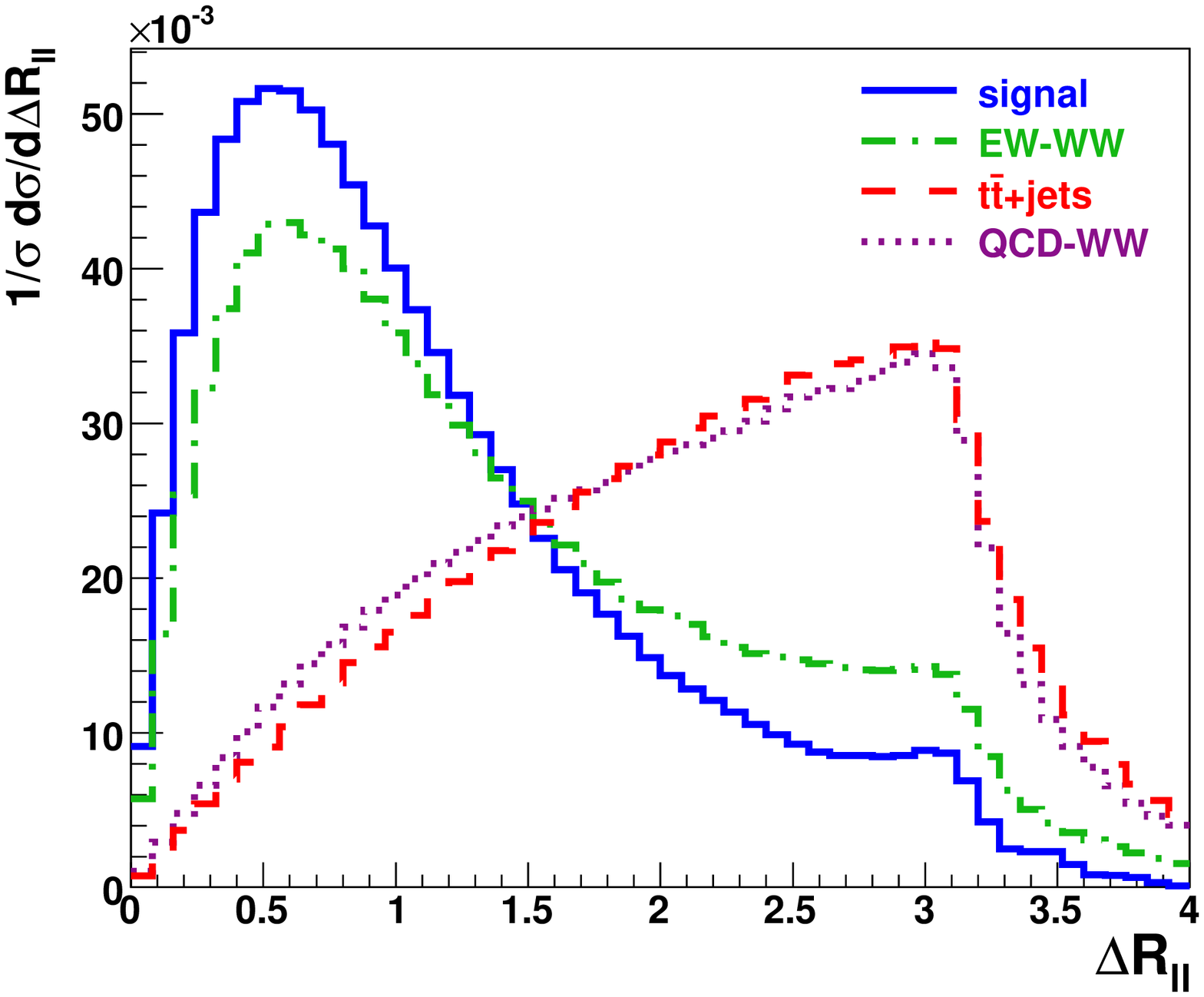}
\includegraphics[scale=0.41]{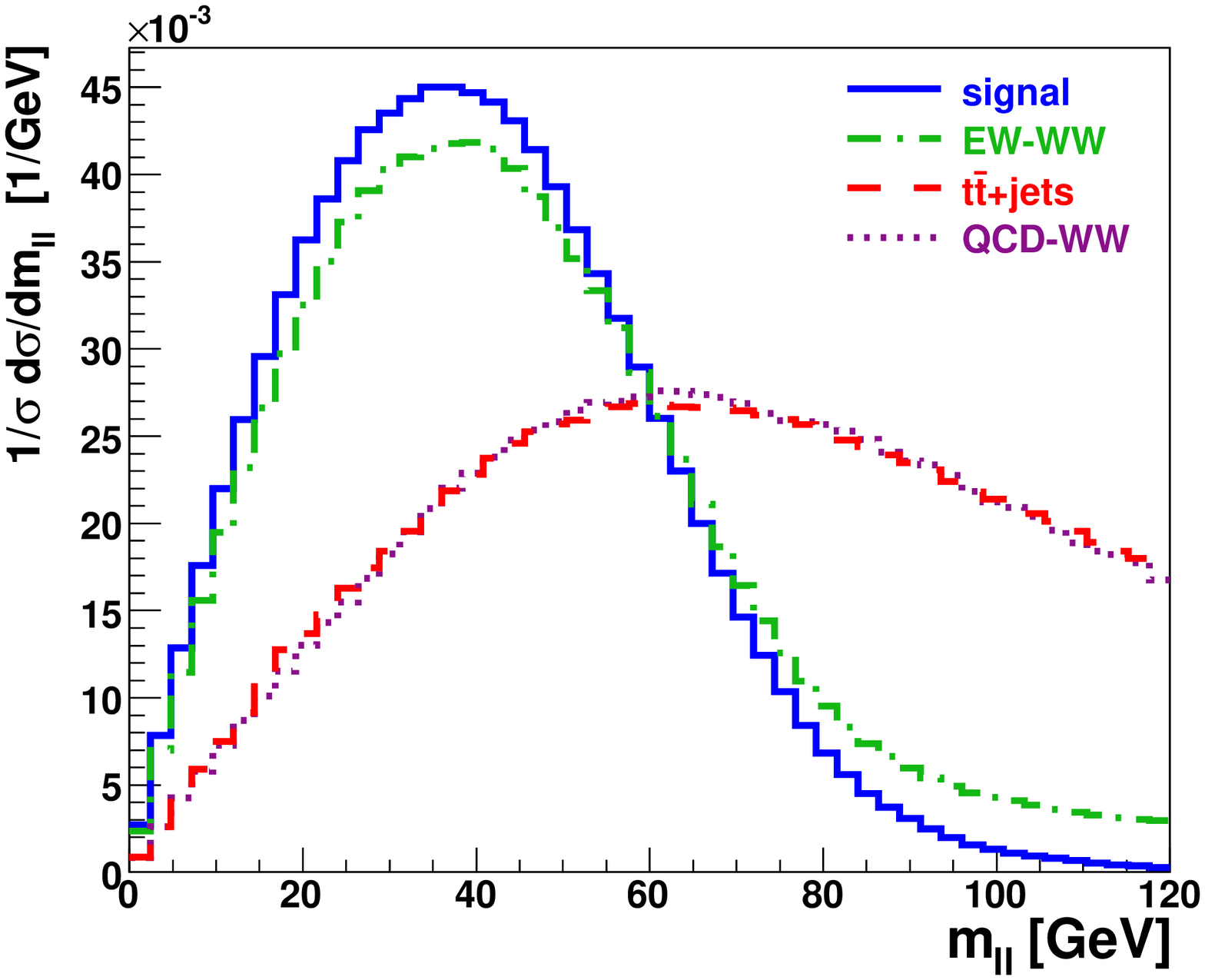}
}
\caption[]{\label{fig:rll} Normalized distributions of the charged
  leptons R-separation {\it (left)} and dilepton invariant mass {\it
  (right)} after cuts of Eq.~(\ref{eqn:incl}) for signal and 
  background as in Fig.~\ref{fig:etajj}.}
\end{figure}

A significant difference between signal and background is provided by
the angle between the charged decay leptons~\cite{Dittmar:1996ss}. In the
case where the $W$ pair is produced via the Higgs decay, the $W$ spins
are anti-correlated, so the leptons are preferentially emitted in the
same direction, close to each other. A large fraction of the backgrounds
does not have anti-correlated $W$ spins. This difference is is
demonstrated in Fig.~\ref{fig:rll} which shows the R-separation
$\Delta R_{\ell\ell}$ and the dilepton invariant mass $m_{\ell\ell}$. The
invariant mass can be expressed by 
\begin{equation}
m_{\ell\ell}=2E_{\ell_1}E_{\ell_2}(1-\cos{\theta_{\ell\ell}})
\end{equation}
with $E_{\ell_{1/2}}$ and $\theta_{\ell\ell}$ being the lepton energy and 
the dilepton opening angle respectively. Hence, a small
opening angle also leads to small $m_{\ell\ell}$ values which is the
case for the Higgs signal as compared to the backgrounds~\cite{Barger:1990mn}.
In Fig.~\ref{fig:rll} the distributions for the electroweak $W^+W^-jj$ background is
very similar to the signal distributions because the EW $W^+W^-jj$
process is dominated by Higgs production. We exploit these
features by imposing the following lepton-pair angular and mass
cuts:
\begin{equation}
\Delta R_{\ell\ell} < 1.1\,,\qquad m_{\ell\ell} < 75\,{\rm GeV}
\label{eqn:mll}
\end{equation}
The results after these cuts are shown on the third line of
Table~\ref{tab:xs}. The signal and the EW $W^+W^-jj$ cross section are
cut in half but the other backgrounds are reduced by roughly one order of
magnitude.
\begin{table}[htb]
  \caption{Signal rates for $m_H=160$ GeV and corresponding background
    cross sections, in $pp$ collisions at $\sqrt{s}=14$ TeV. Results are
    given for various levels of cuts and are labeled by equation numbers
    discussed in the text. All rates are given in fb. The last two
    columns give the signal to background ratio $S/B$ and the
    $S/\sqrt{B}$ ratio for an assumed integrated luminosity of $30\,{\rm
    fb}^{-1}$.}
\begin{center}
\renewcommand{\arraystretch}{1.15}
\begin{tabular}{|l|cccccc|cc|}
\hline
& & EW  & & & & QCD & & \\
cuts & GF & $WWjj$  & $t\bar{t}$ & $t\bar{t}j$ & $t\bar{t}jj$ &
$WWjj$ & $S/B$ & $S/\sqrt{B}$\\
\hline
inclusive cuts (\ref{eqn:incl}) & 
     115.2 & 75.1 & 6832 & 9518 & 1676 & 363 & 1/160 & 4.6\\
+ $b$ veto & 
      99.2 & 67.4 &  833 & 1822 &  564 & 307 & 1/36 & 9.1 \\
+ $R_{\ell\ell}$, $m_{\ell\ell}$ cut (\ref{eqn:mll})   &  
      55.8 & 30.7 &  104 &  218 & 86.4 & 42.7 & 1/8.6 & 13.9\\
+ $p_{T\ell}$ cut (\ref{eqn:ptl})      &  
      41.5 & 22.3 & 38.3 & 87.7 & 29.2 & 20.5 & 1/4.8 & 16.2\\ 
+ $m_T^{WW}$ cuts (\ref{eqn:mtww},\ref{eqn:mllmt})   & 
      37.1 & 19.9 & 30.1 & 63.4 & 19.3 & 13.4 & 1/3.8 & 16.8\\
+ $\fdag{p}_T$ cut (\ref{eqn:ptmiss})  &
      31.5 & 16.5 & 23.3 & 51.1 & 11.2 & 11.4 & 1/3.6 & 16.2\\
\hline
\end{tabular}
\end{center}
\label{tab:xs}
\end{table} 

Examining the lepton $p_T$ distributions, it turns out that 
background events which survive the angular cut of Eq.~(\ref{eqn:mll})
have a significantly lower lepton $p_T$ than the signal and VBF
events.  This is demonstrated in Fig.~\ref{fig:ptlmin}: 
\begin{figure}[htb]
\centerline{
\includegraphics[scale=0.41]{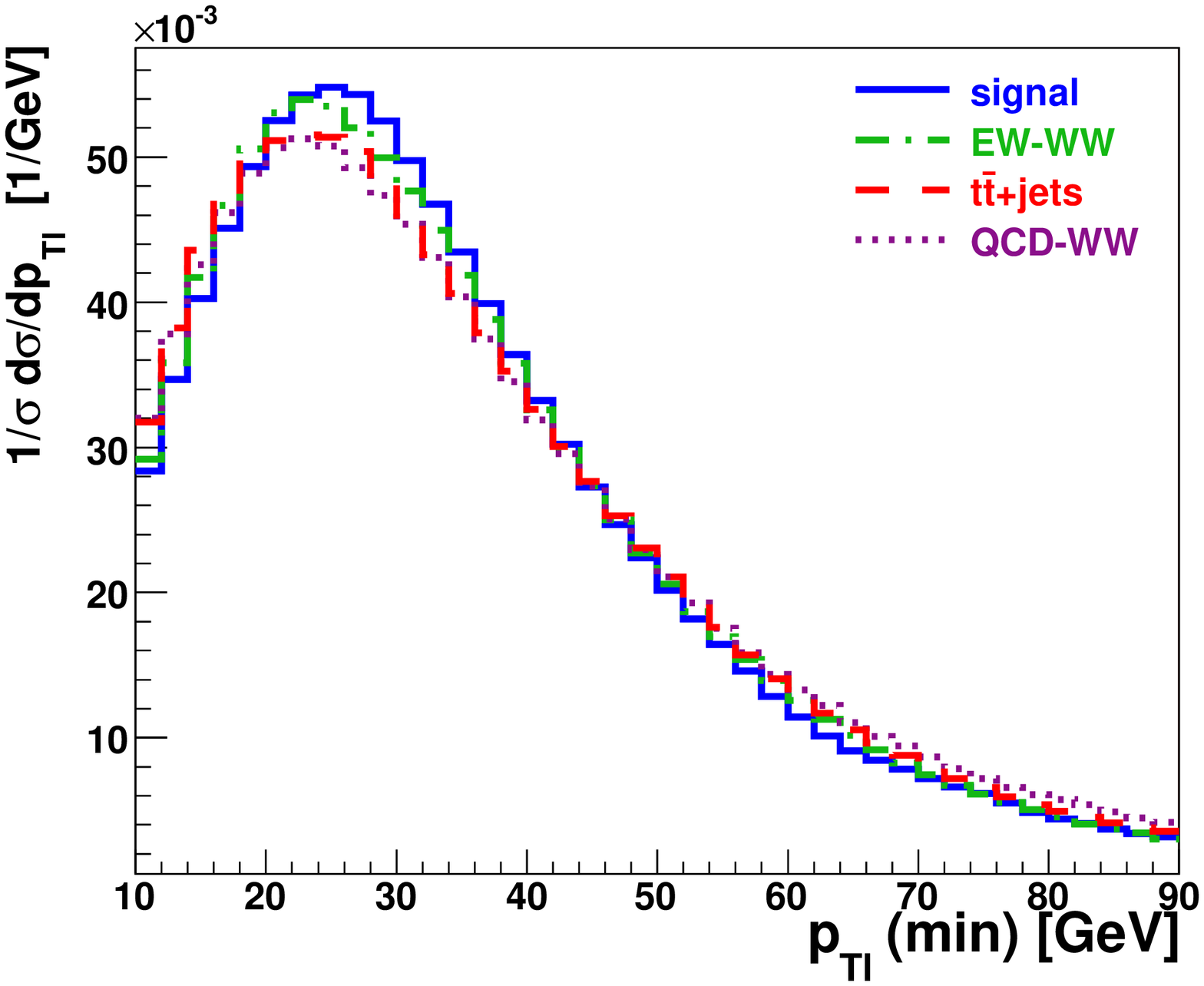}
\includegraphics[scale=0.41]{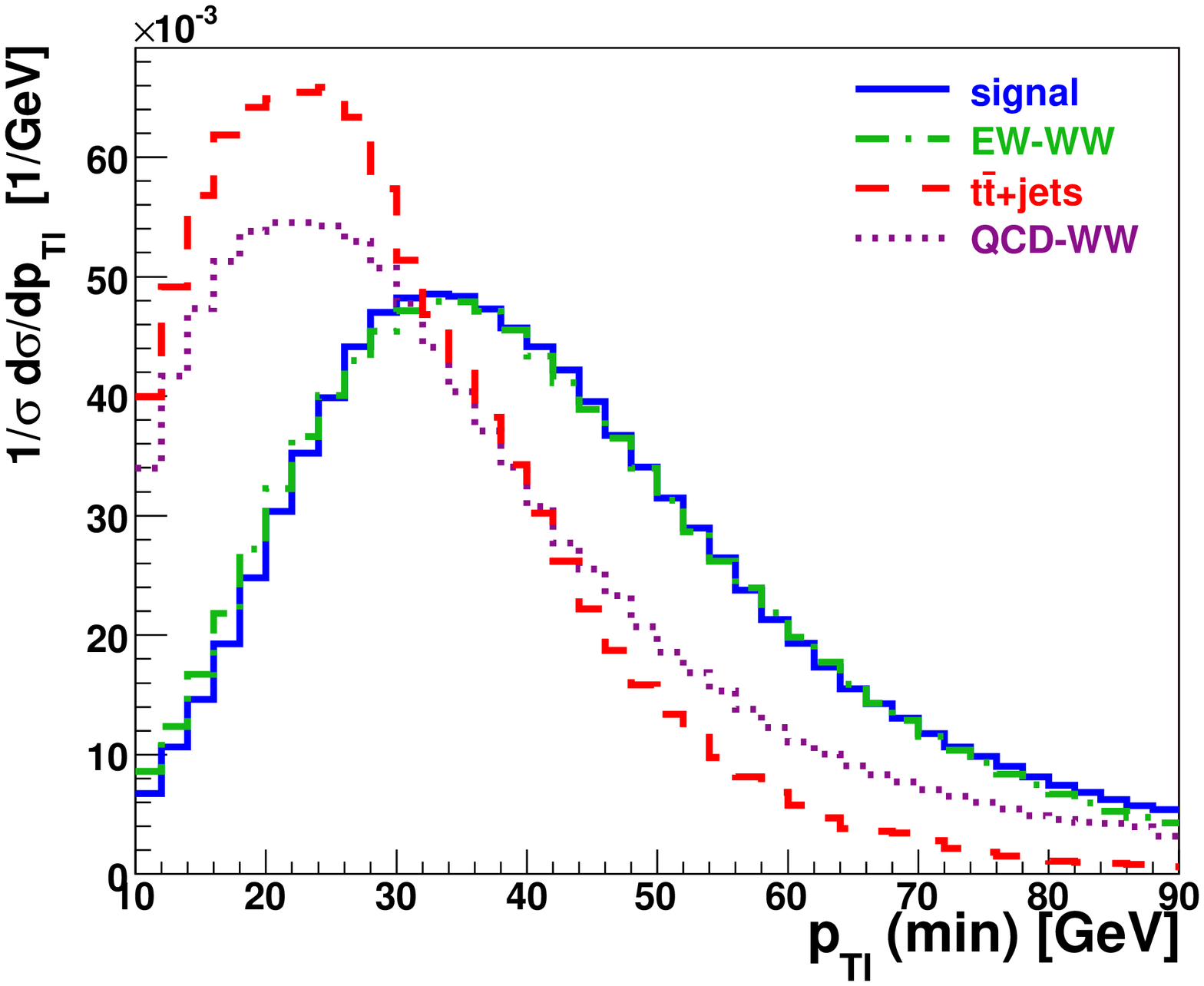}
}
\caption[]{\label{fig:ptlmin} Normalized distributions of the minimum
  charged lepton $p_T$ for inclusive cuts of Eq.~(\ref{eqn:incl}) {\it
  (left)} and after the additional $\Delta R_{\ell\ell}$ and
  $m_{\ell\ell}$ cuts of
  Eq.~(\ref{eqn:mll}) {\it (right)}. Curves are for signal and
  backgrounds as in Fig.~\ref{fig:etajj}. }
\end{figure}
On the left hand side the distributions of minimum lepton $p_T$ are plotted
for signal and background, for the inclusive cuts Eq.~(\ref{eqn:incl}). The
curves lie almost on top of each other and there seems to be no
difference between the various processes. The right hand side of
Fig.~\ref{fig:ptlmin} shows the same distributions after applying the
$R_{\ell\ell}$ and $m_{\ell\ell}$ cut. This suggests a harder lepton
$p_T$ cut and we impose 
\begin{equation}
P_{T\ell} > 30\,{\rm GeV}
\label{eqn:ptl}
\end{equation}
in the following. Note that this harder cut on the two charged leptons
implies excellent trigger efficiencies even in high luminosity running.

\begin{figure}[hbt]
\centerline{
\includegraphics[scale=0.41]{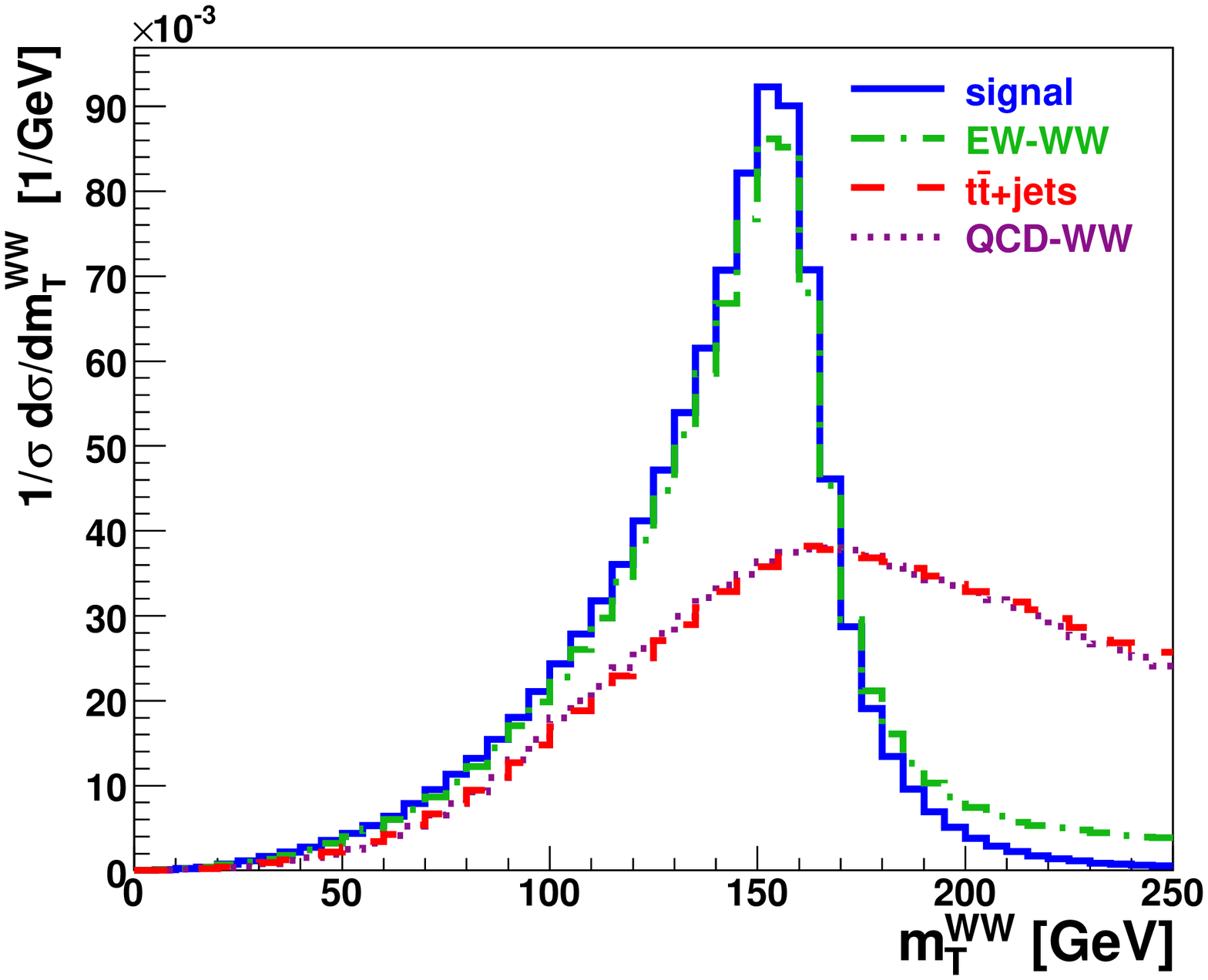}
\includegraphics[scale=0.41]{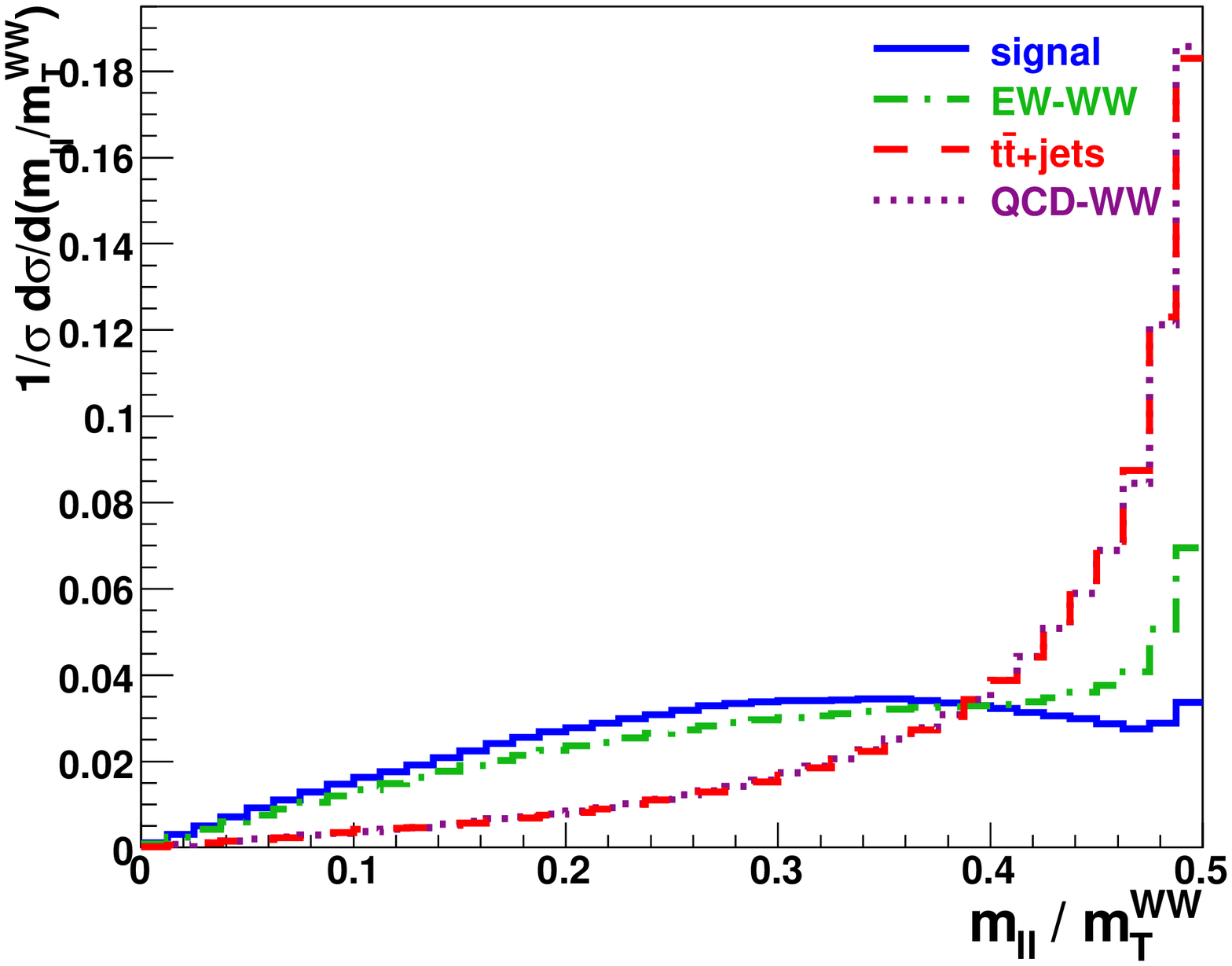}
}
\caption[]{\label{fig:mtww} {\it Left:} Normalized distribution of the
transverse mass of the dilepton system as defined in
Eq.~(\ref{eqn:mtdef}). {\it Right:} The ratio of the dilepton invariant
mass $m_{\ell\ell}$ and transverse mass $m_T^{WW}$. Curves are for
signal and backgrounds as in Fig.~\ref{fig:etajj} with the inclusive cuts of Eq.~(\ref{eqn:incl}).}
\end{figure}

It is known that the transverse mass of the dilepton-$\fdag{\vec{p}}_T$ system
can be used to reconstruct the Higgs boson mass. This works particularly well
for masses at or below $W$ pair threshold. We
here use the transverse mass definition of Ref.~\cite{VBF:H},
\begin{equation}
m_T^{WW} = \sqrt{(\fdag{E}_T+E_{T,{\ell\ell}})^2-
({\vec{p}}_{T,{\ell\ell}}+\fdag{\vec{p}}_T)^2}
\label{eqn:mtdef}
\end{equation}
in terms of the invariant mass of the two charged lepton and the 
transverse energies 
\begin{equation}
E_{T,\ell\ell} = (p_{T,\ell\ell}^2 + m_{\ell\ell}^2)^{1/2},\qquad
\fdag{E}_T = (\fdag{p}_T^2 + m_{\ell\ell}^2)^{1/2} .
\label{eqn:etdef}
\end{equation}
In Fig.~\ref{fig:mtww} the GF signal and the VBF process show a
pronounced Jacobian peak in the $m_T^{WW}$ distribution whereas the
$t\bar{t}+jets$ and QCD $W^+W^-jj$ backgrounds events are broadly
distributed. 
Since the process of gluon induced Higgs$+2$~jet production will not be
a discovery channel for the Higgs boson, we can assume that the Higgs
boson mass is known and we can further optimize our cuts for this
mass. Hence we impose a strict cut on the $m_T^{WW}$ observable:
\begin{equation}
m^{WW}_T < 170\,{\rm GeV}
\label{eqn:mtww}
\end{equation}
Considering Eqs.~(\ref{eqn:mtdef}) and (\ref{eqn:etdef}), $m_T^{WW}$ and
$m_{\ell\ell}$ are apparently correlated with each other. We find
that the ratio $m_{\ell\ell}/m^{WW}_T$ contains further useful information.
We apply a cut of the form:
\begin{equation}
m_{\ell\ell}< 0.44 \cdot m^{WW}_T 
\label{eqn:mllmt}
\end{equation}
With these cuts the backgrounds are again strongly reduced while the Higgs
induced processes are affected at the 10\% level only. 
The results for the cuts of Eqs.~(\ref{eqn:ptl}), (\ref{eqn:mtww}) and
(\ref{eqn:mllmt}) are shown in line 4 and 5 of Table~\ref{tab:xs}.

\begin{table}[tbh]
  \caption{Signal and background cross sections and the expected
  number of events for ${\cal L}_{int}=30\,{\rm fb}^{-1}$ at different
  levels of cuts.}
\begin{center}
\begin{tabular}{|c|c|c|c|c|c|}
 \cline{2-6}
\multicolumn{1}{c|}{} & inclusive cuts & \multicolumn{2}{c|}{selection
  cuts} & \multicolumn{2}{c|}{selection
  cuts + Eq.~(\ref{eqn:eta}) }\\
\hline
process & $\sigma$ [fb] & $\sigma$ [fb] & events / {$30\,{\rm fb}^{-1}$}
&  $\sigma$ [fb] & events / {$30\,{\rm fb}^{-1}$}\\
\hline\hline
GF $pp \rightarrow H +j j$       & 115 & 31.5 & 945 & 10.6 & 318 \\
\hline
EW $pp \rightarrow W^+W^- +j j$ & 75 & 16.5 & 495 & 13.9 & 417 \\
$pp \rightarrow t \bar{t}$       & 6830 & 23.3 & 699 & 1.5 & 45 \\
$pp \rightarrow t \bar{t}+ j$    & 9520 & 51.1 & 1530 & 13.4 & 402 \\
$pp \rightarrow t \bar{t}+ jj $  & 1680 & 11.2 &  336 & 3.8 & 114 \\
QCD $pp \rightarrow W^+W^- +j j$ & 363  & 11.4 & 342 & 3.0 & 90 \\
\hline
sum of backgrounds             & 18500 & 114  & 3410 & 35.6 & 1070 \\
\hline
\end{tabular}
\end{center}
\label{tab:xs2}
\end{table}

\begin{figure}[tbh]
\centerline{
\includegraphics[scale=0.45]{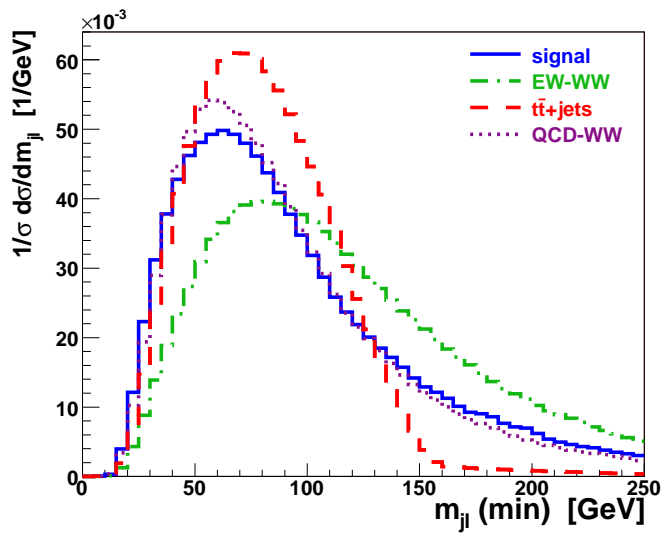}
}
\caption[]{\label{fig:mjl} Normalized distributions for the
  minimum jet-lepton invariant mass. Curves are for signal and backgrounds as in Fig.~\ref{fig:etajj}
  with the inclusive cuts of Eq.~(\ref{eqn:incl}).
}
\end{figure}

At this level the signal rate is reduced by a factor of 3 as compared to the
inclusive cuts, but the backgrounds now have cross sections of the same
order as the signal. The largest background still arises from the
$t\bar{t}$ processes, especially $t\bar{t}+1j$, i.e. one tagging jet
arises from an (unidentified) $b$-quark from $t$ or $\bar{t}$ decay and
the other one is due to emission of a light quark or gluon in $t\bar{t}$
production. For an integrated
luminosity of ${\cal L}_{int}=30\,{\rm fb}^{-1}$ the rates correspond to a
purely statistical significance of the gluon fusion signal of
$S/\sqrt{B}\approx 17$. 
However, additional backgrounds arise from $\ell\ell jj$ events
where the missing transverse momentum is generated by detector
effects. It has been shown in Ref.~\cite{VBF:H} that these backgrounds
are under control when requiring a missing $p_T$ of at least 30
GeV:
\begin{equation}
\fdag{p}_T > 30\,{\rm GeV}
\label{eqn:ptmiss}
\end{equation}
The $\fdag{p}_T$ distributions are shown on the right
hand side of Fig.~\ref{fig:pt}. The GF signal and the backgrounds which
we have considered are affected
similarly by this cut. The resulting cross sections are shown in the
last line of Table~\ref{tab:xs}. We are left with a signal to background
ratio of one over $3.6$. For ${\cal L}_{int}=30\,{\rm fb}^{-1}$ we expect
945 signal events on top of 3400 background events as summarized in
Table~\ref{tab:xs2}. 
This corresponds to a statistical significance of $S/\sqrt{B}\approx
16$. Note, however, that the top-production backgrounds need to be
understood with an accuracy of 7\% or better in the signal region in
order to allow for a $5\sigma$ Higgs signal from rate measurements
alone. If background uncertainties turn out to be that large one may
want to reexamine the selection. A significantly higher
signal to background ratio, and a concomitant smaller effect of
background uncertainties,  can be achieved by more aggressive cuts, as is
obvious from the various distributions in Figs.~\ref{fig:rll} to
\ref{fig:mjl}. In the absence of solid estimates for systematic errors
we have tried to optimize the statistical significance
and leave further refinements to future studies. In any case, the event
rates summarized in Table~\ref{tab:xs2} are sufficiently large 
to allow the analysis of distributions.

It is clear that a cut based analysis as described above is not
the optimal way to isolate the signal from the
background. More advanced techniques like neural networks or
likelihood methods should be considered. As input for such refinements
we show one further observable,
which, however, could not be exploited in our cut based approach.
The minimum jet-lepton transverse mass $m_{jl}$, i.e. the minimum of the four
combinations of the charged leptons with the two tagging jets, is
plotted in Fig.~\ref{fig:mjl}. The distribution shows that $m_{jl}$ is
bounded by the top mass for the $t\bar{t}$ processes. Only a small
tail is left above $m_t$ for the $t\bar{t}jj$ process. This characteristic
remains after the selection cuts. Since the contribution from top
backgrounds is so different for $m_{jl}\lsim 150$ GeV and $m_{jl}\gsim
150$ GeV, it might be useful to perform separate cut optimizations for
the two regions.

\section{Azimuthal angle correlations}
\label{sec:azimjj}

In order to determine the tensor structure of the effective $Hgg$
coupling, the distributions of the two tagging jets are an important
tool. The distribution $d\sigma/d|\Delta\Phi_{jj}|$ of the azimuthal
angle between the two tagging jets provides for an excellent 
distinction between the two tensor structures of 
Eq.~(\ref{eq:Tmunu})~\cite{VBF:CP}. Unfortunately, when both
CP-even and CP-odd couplings of similar strength are present, the
tensor structure cannot be unambiguously determined anymore.
The missing information is contained in the sign of the azimuthal
angle between the tagging jets~\cite{Hankele:2006ma}. 
Naively one might assume that this
sign cannot be defined unambiguously in $pp$ collisions because an
azimuthal angle switches sign when viewed along the opposite beam
direction. However, in doing so, the ``toward'' and the ``away''
tagging jets also switch place, i.e. one should take into account the
correlation of the tagging jets with the two distinct beam
directions. Defining $\Delta\Phi_{jj}$ as the azimuthal angle of the
``away'' jet minus the azimuthal angle of the ``toward'' jet, a switch
of the two beam directions leaves the sign of $\Delta\Phi_{jj}$
intact. 
\begin{figure}[hbt]
\centerline{
\includegraphics[scale=0.55]{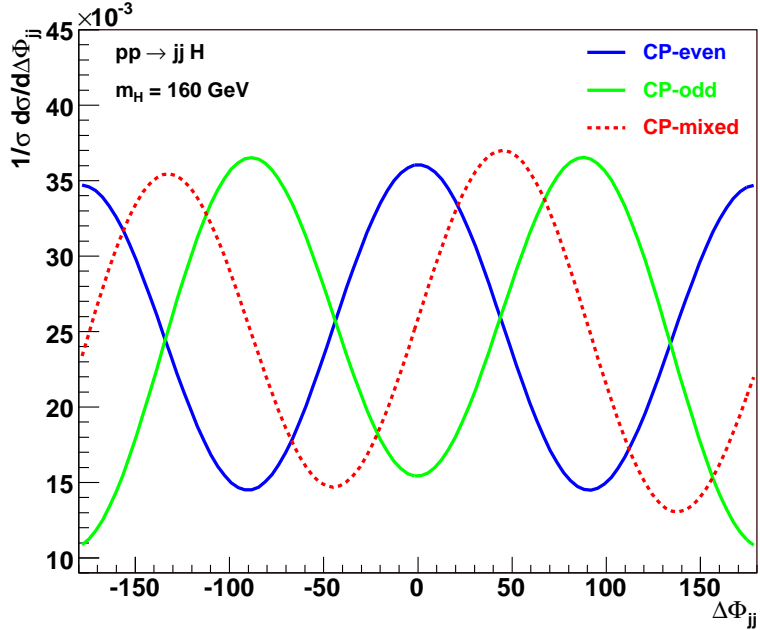}
}
\caption[]{\label{fig:phi1} Normalized distributions of the jet-jet
  azimuthal angle difference as defined in Eq.~(\ref{eq:phijj}). The
  curves are for the SM CP-even case ($a_{3}=0$), a pure CP-odd
  ($a_{2}=0$) and a CP-mixed case ($a_{2}=a_{3}\ne 0$). The cuts of
  Eq.~(\ref{eqn:incl}) and (\ref{eqn:eta}) were applied. 
}
\end{figure}
To be precise, let us define the normalized
four-momenta of the two proton beams as $b_+$ and $b_-$, while $p_+$
and $p_-$ denote the four momenta of the two tagging jets, where $p_+$
points into the same detector hemisphere as $b_+$. Then
\begin{equation}
\label{eq:phijj}
\varepsilon_{\mu\nu\rho\sigma} b_+^\mu p_+^\nu b_-^\rho p_-^\sigma
= 2p_{T,+}p_{T,-}\sin(\phi_+ - \phi_-) = 2p_{T,+}p_{T,-}\sin\Delta\Phi_{jj}
\end{equation}
provides the sign of $\Delta\Phi_{jj}$. This definition is manifestly
invariant under the interchange $(b_+,p_+)\leftrightarrow (b_-,p_-)$,
i.e. when viewing the event from the opposite beam direction,  and
we also note that $\Delta\Phi_{jj}$ is a parity odd observable.

The corresponding azimuthal angle distribution is shown in
Fig.~\ref{fig:phi1} for the gluon fusion Higgs signal 
for three scenarios of CP-even and CP-odd Higgs
couplings. All three cases are well distinguishable.
The maxima in the distribution are directly connected to the size
of the parameters $a_2$ and $a_3$, which were introduced in 
Eqs.~(\ref{eq:Tmunu},\ref{eq:a2a3}). For
\begin{equation}
a_2 = a\,\cos\alpha\, , \qquad a_3 = a\, \sin \alpha\, ,
\end{equation}
the positions of the maxima are at $\Delta\Phi_{jj}=\alpha$ and
$\Delta\Phi_{jj}=\alpha \pm \pi$. This also explains why
$|\Delta\Phi_{jj}|$ loses information in the mixed CP case: when
folding over the $\Delta\Phi_{jj}$-distribution at
$\Delta\Phi_{jj}=0$, maxima and minima at $+45$ and $-45$ degrees 
cancel each other.

As noted above, $\Delta\Phi_{jj}$ is a parity odd observable. Finding a
$\Delta\Phi_{jj}$ asymmetry as in Fig.~\ref{fig:phi1} would show
that parity is violated in the process $pp\to HjjX$. Since the QCD
couplings are parity conserving, the parity
violation must originate from a parity-odd Higgs coupling, namely $a_3$
in the effective $Hgg$ vertex. This term is also CP-odd. Such a coupling, occurring
at the same time as the CP-even SM coupling $a_2$, implies CP-violation in the
Higgs sector. In this sense, the observation of an asymmetry in the
$\Delta\Phi_{jj}$ distribution would directly demonstrate CP-violation
in the Higgs sector.
\begin{figure}[tbh]
\centerline{
\includegraphics[scale=0.41]{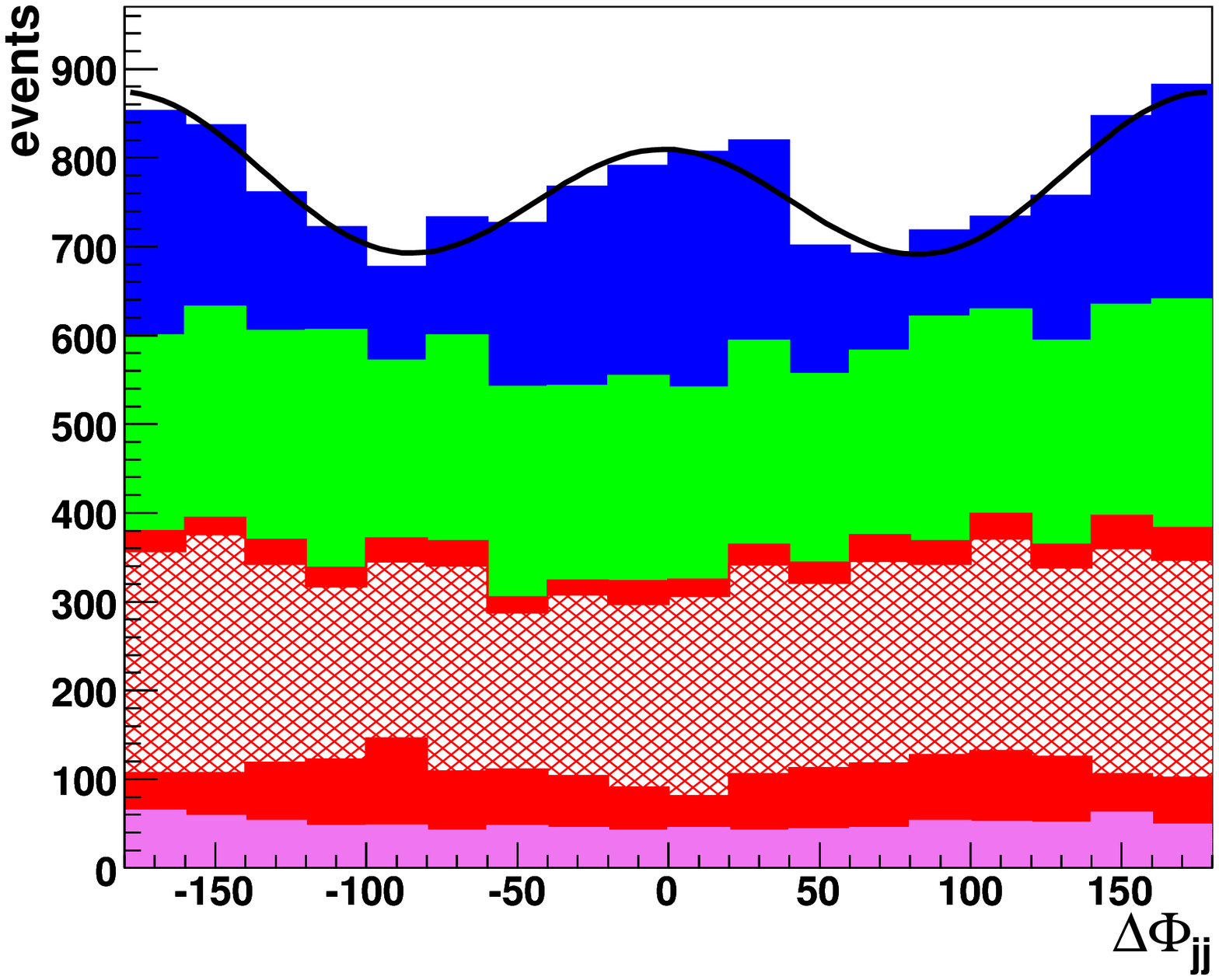}
\includegraphics[scale=0.41]{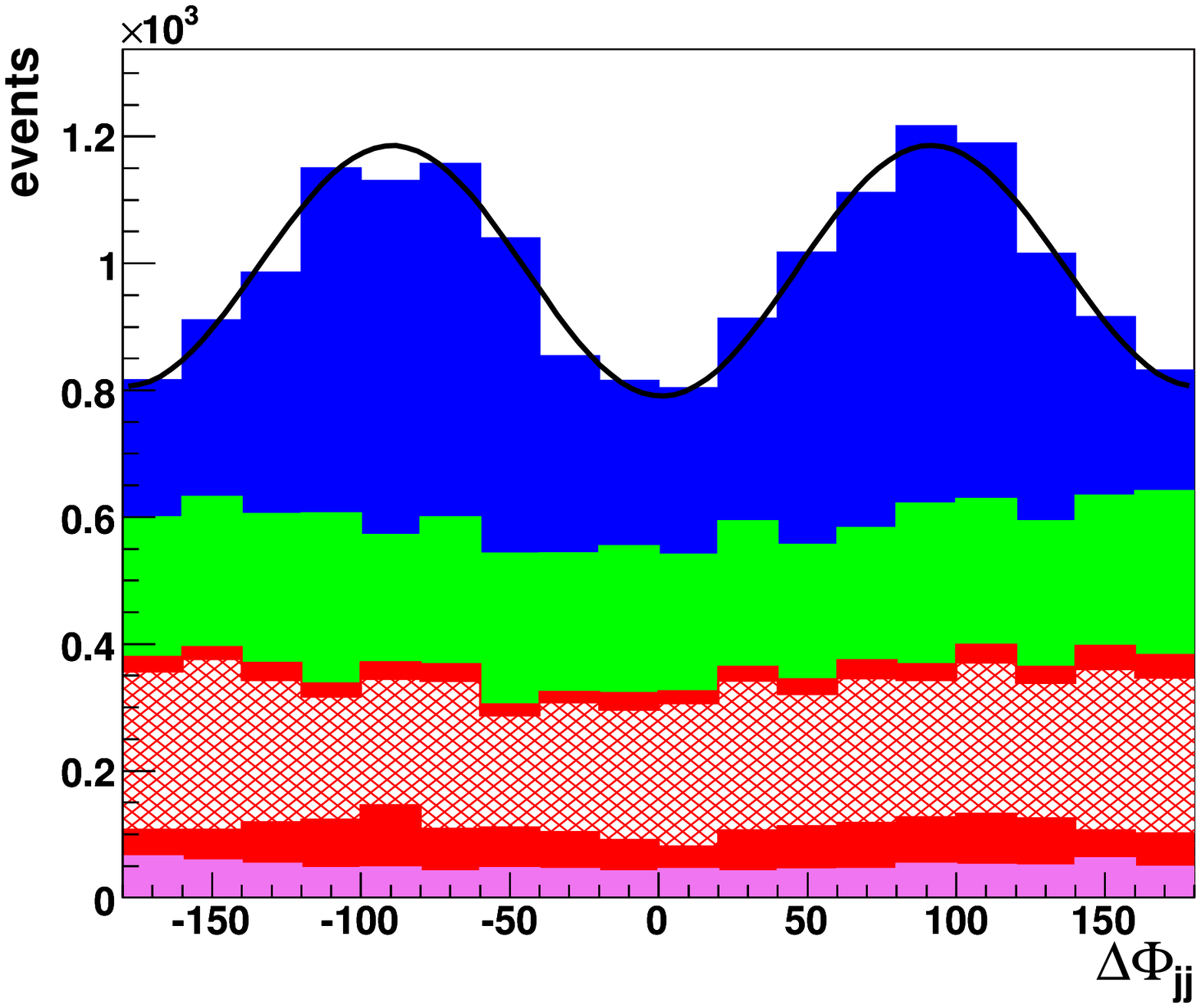}
}
\caption[]{\label{fig:phijjfit} The $\Delta\Phi_{jj}$ distribution for
a pure CP-even coupling {\it (left)} and a pure CP-odd coupling {\it
  (right)} for ${\cal L}_{int} = 300\,{\rm fb}^{-1}$. From top to
bottom: GF signal, EW $W^+W^-jj$, $t\bar{t}$,
$t\bar{t}j$, $t\bar{t}jj$, and QCD $W^+W^-jj$ backgrounds. }
\end{figure}

The azimuthal angle difference $\Delta\Phi_{jj}$ of the two tagging jets
is the observable that carries information about the CP nature of the
$Htt$ coupling. After we have improved the $S/\sqrt{B}$ ratio as
much as possible by means of selection cuts, we now want to extract this
information from the $\Delta\Phi_{jj}$ distribution. For this purpose we
define a fit function
\begin{equation}
f(\Delta\Phi) = N(1 + A\cos[2(\Delta\Phi-\Delta\Phi_{max})]-B\cos(\Delta\Phi) ),
\label{eqn:fit}
\end{equation}
with fit-parameters $A, \Delta\Phi_{max}, B, N$. The parameter $N$ is a
normalization factor and $B$ is an overall shape-factor. The parameters
$A$ and $\Delta\Phi_{max}$ are the physically relevant ones. $A$ describes
the relative magnitude of the angle correlation and thus the
significance of this measurement with respect to the background
fluctuations. $\Delta\Phi_{max}$ gives the position of the
first maximum, which measures the relative strength of CP-even and
CP-odd couplings.

\begin{figure}[bht]
\centerline{
\includegraphics[scale=0.41]{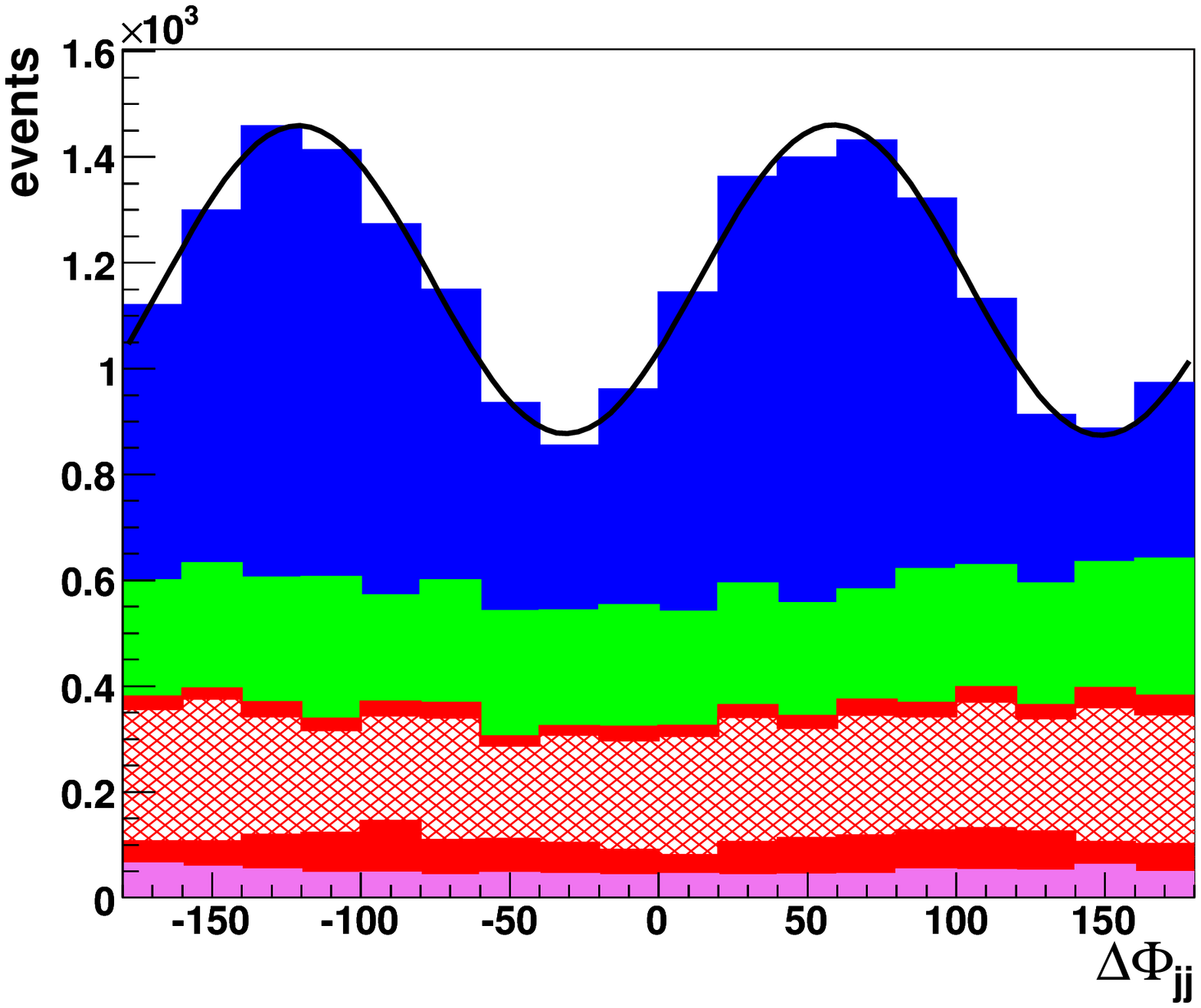}
\includegraphics[scale=0.41]{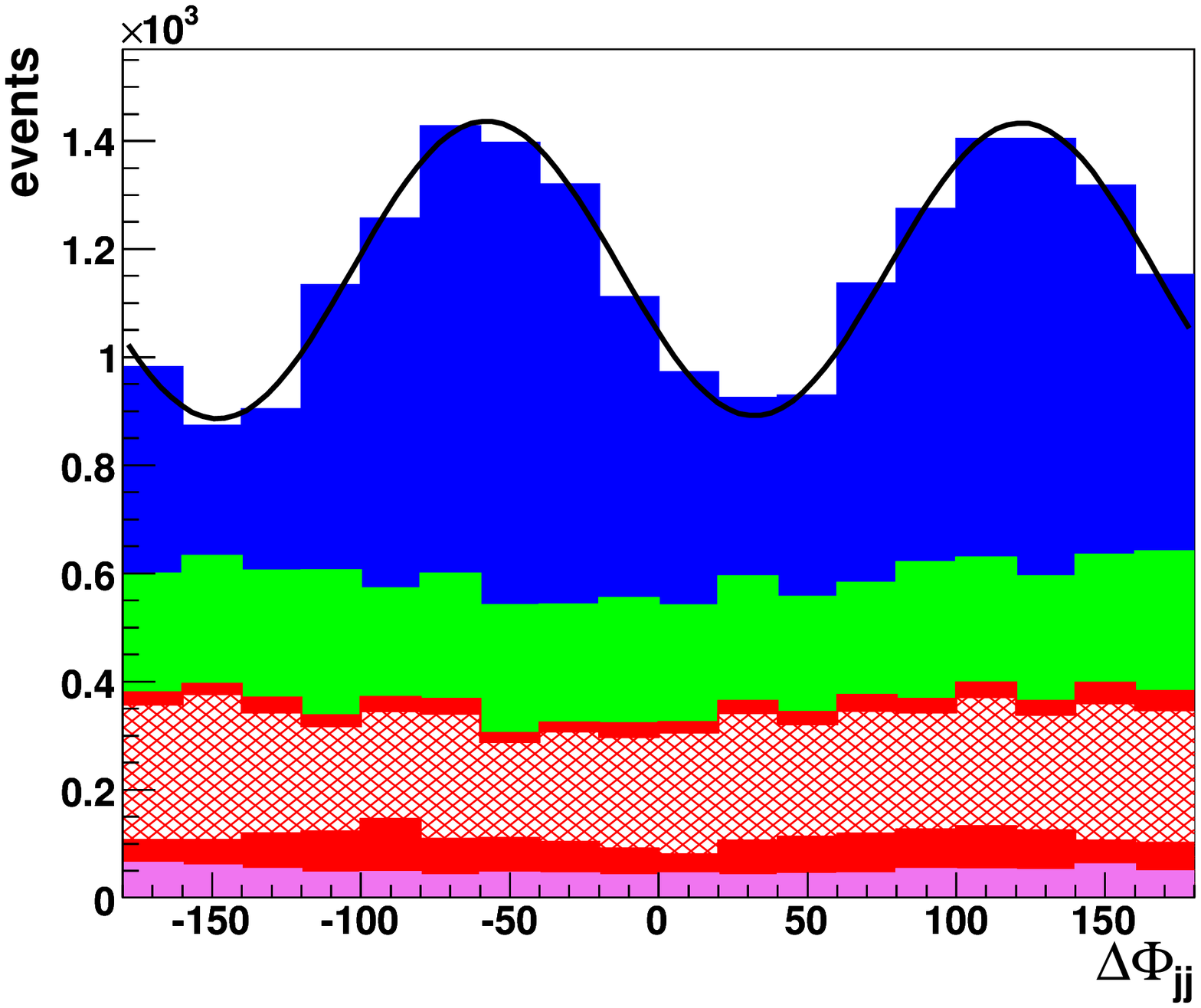}
}
\caption[]{\label{fig:phijjfit2} The $\Delta\Phi_{jj}$ distribution
  for CP-mixed couplings as in Fig.~\ref{fig:phijjfit} for 
  ${\cal L}_{int} = 300\,{\rm fb}^{-1}$. {\it Left:} 
  $y_t=\tilde y_t = y_t^{SM}$. {\it Right:}
  $y_t=-\tilde y_t = y_t^{SM}$. }
\end{figure}

 Figure~\ref{fig:phijjfit} shows the expected $\Delta\Phi_{jj}$
distribution for a purely CP-even ($y_t=y_t^{SM}$, $\tilde y_t=0$) and a
purely CP-odd $Htt$ coupling ($y_t=0$, $\tilde y_t =y_t^{SM}$) for an
integrated luminosity of $300\,{\rm fb}^{-1}$.  
Plotted are signal events on top of the various
backgrounds. The black curve is the fit to this distribution with the
function of Eq.~(\ref{eqn:fit}). A comparison with
Fig.~\ref{fig:phi1} shows that the relevant characteristic angular correlation
is kept but diluted due to the background. However with the help of the
fit we can extract the parameters $A$ and $\Delta\Phi_{max}$. In order
to estimate the statistical significance of the measurement, we divide
our Monte Carlo sample into 10 independent samples of data,
each corresponding to an integrated luminosity of $30\,{\rm fb}^{-1}$.
Averaging the results and the errors, we obtain $A=0.115\pm0.039$,
$\Delta\Phi_{max}=-3.0\pm 10.7$ for the CP-even coupling and
$A=0.210\pm0.034$, $\Delta\Phi_{max}=90.4\pm4.7$ for the CP-odd
coupling, where errors are purely
statistical. The expected values would be $\Delta\Phi_{max}=0$ and
$\Delta\Phi_{max}=90$ respectively. Since the difference between the
CP-even and the CP-odd case can also be expressed as a flip in the sign
of $A$ (keeping $\Delta\Phi_{max}=0$) this result means that a CP-odd
$Htt$ coupling can be distinguished from the SM case with approximately
$6\sigma$ significance for 30~fb$^{-1}$, assuming a perfect detector.
Since the backgrounds are fairly flat in $\Delta\Phi_{jj}$, any
background normalization uncertainties are largely absorbed into the fit
parameters $N$ and $B$. Our estimates for the significance of the $A$
determination will therefore receive minor changes due to
such systematic errors.

Fig.~\ref{fig:phijjfit2} shows the the azimuthal angle distribution
for the case of a CP-mixed coupling $y_t=\pm\tilde y_t = y_t^{SM}$. 
Applying the method described above, we extract $A=0.260\pm0.031$,
$\Delta\Phi_{max}=59.3\pm 3.5$ from the left part of
Fig.~\ref{fig:phijjfit2} and $A=0.246\pm0.031$,
$\Delta\Phi_{max}=-58.4\pm 3.7$ from the right one. The expected
values would be $\Delta\Phi_{max}=\pm\arctan(\frac{3}{2})=\pm 56.3$.
\begin{figure}[bht]
\centerline{
\includegraphics[scale=0.41]{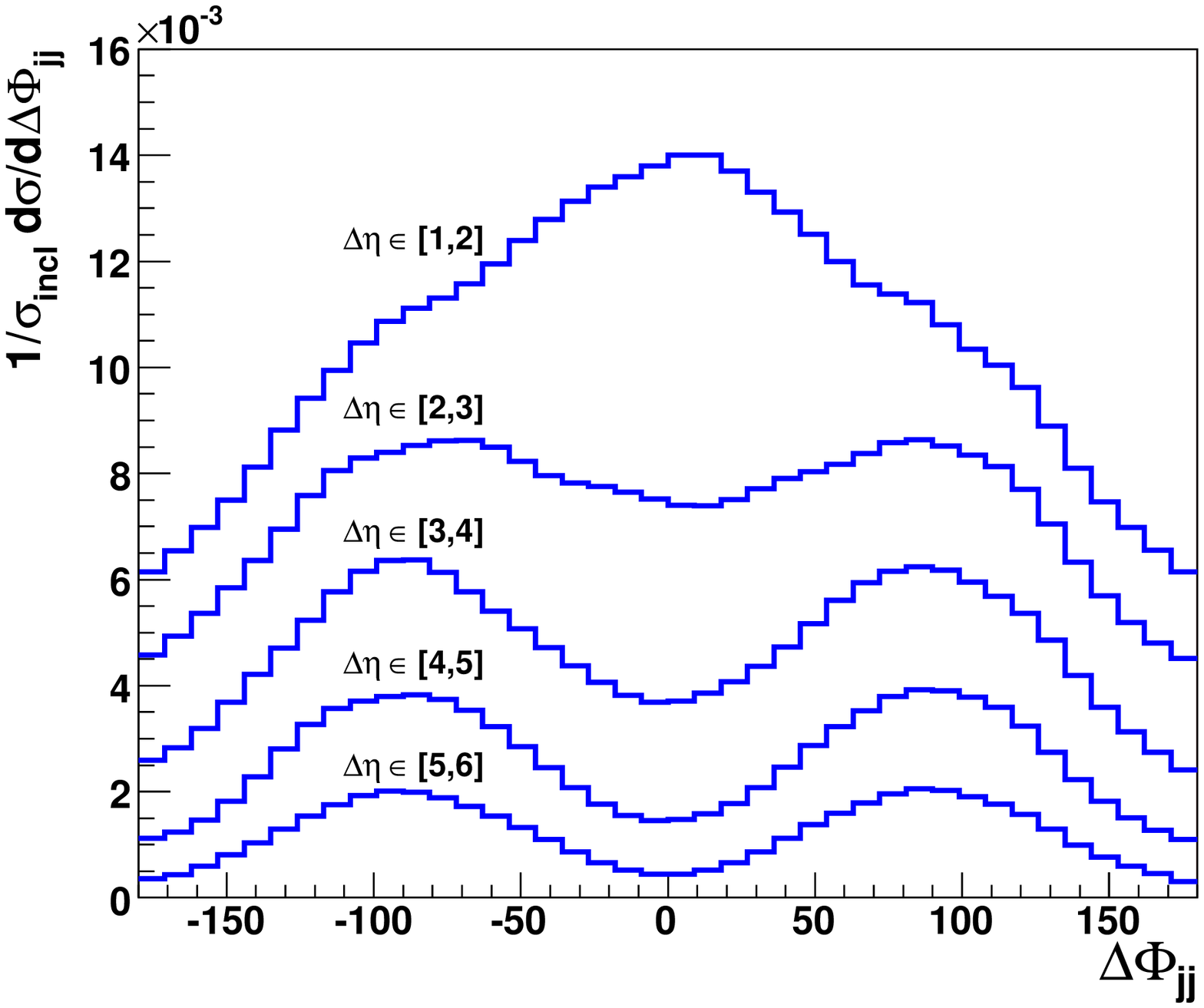}
\includegraphics[scale=0.41]{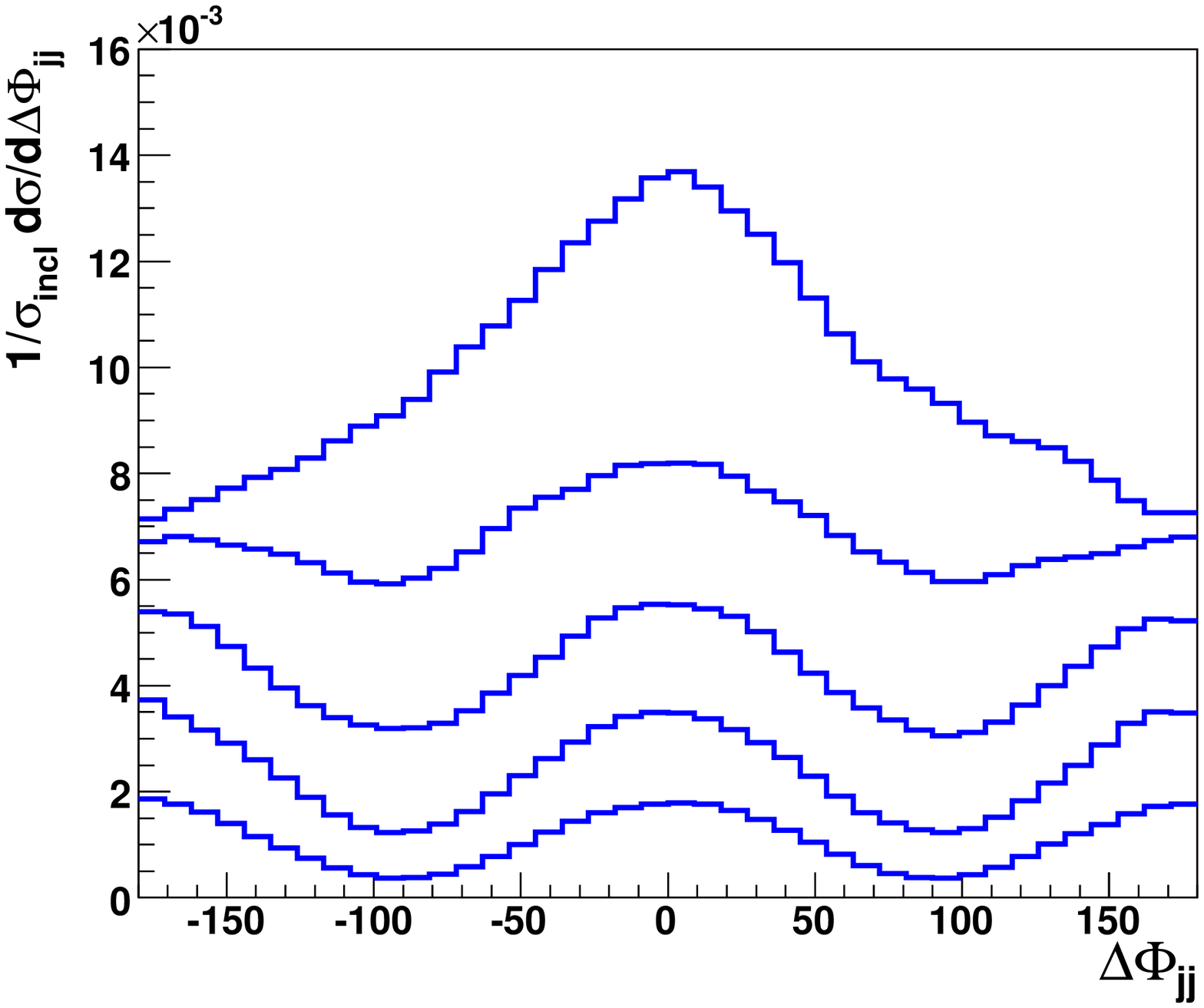}
}
\caption[]{\label{fig:phijjeta1} The $\Delta\Phi_{jj}$ distribution of
  the signal for different intervals of the jet rapidity separation
  $\Delta\eta_{jj}$ as labeled in the left plot. {\it Left:} CP-even
  coupling. {\it Right:} CP-odd coupling. }
\end{figure}

For the procedure described above, we have applied an additional cut on
the dijet rapidity separation of 
\begin{equation}
\Delta\eta_{jj}=|\eta_{j_1}-\eta_{j_2}| > 3.0
\label{eqn:eta}
\end{equation}
This is advantageous because the analyzing power of the
$\Delta\Phi_{jj}$ distribution strongly
depends on the rapidity separation of the two tagging jets. This is demonstrated in
Fig.~\ref{fig:phijjeta1} which shows the distributions within different
$\Delta\eta_{jj}$ intervals for the CP-even and the CP-odd case. For low
$\Delta\eta_{jj}$ values the $\Delta\Phi_{jj}$ distribution does not
show the characteristic features discussed above. However, they get more
pronounced as $\Delta\eta_{jj}$ increases.
Therefore, a cut on $\Delta\eta_{jj}$ raises the analyzing power of the
azimuthal angle distribution. Unfortunately, an additional cut also 
decreases the number of signal events and thus the significance of 
the measurement. It turns out that a cut value around
$\Delta\eta_{cut}\approx 3$ leads to a minimal error on
$\Delta\Phi_{max}$ and an optimal $\Delta A/A$ ratio.

Our analysis has been performed with parton level Monte Carlo programs
based on leading order matrix elements for the signal and the
backgrounds. In the similar situation of widely separated dijets in
$p\bar p$ collisions at the Tevatron, some de-correlation in the
azimuthal angle distribution is predicted at higher order at the parton 
level~\cite{DelDuca:1994mn,Stirling:1994zs,DelDuca:1995fx,DelDuca:1995ng,Orr:1997im}
and with parton showers and 
hadronisation~\cite{Marchesini:1988cf,Marchesini:1992ch}, and measured
at the Tevatron~\cite{Abachi:1996et}. A strong de-correlation, as
originally predicted in a pure parton shower approach in
Ref.~\cite{Odagiri:2002nd}, would invalidate our method for Higgs CP
studies. Using full leading order QCD matrix elements for $H+2$- and
$H+3$-parton production and additional parton shower simulation, 
the question was reanalyzed recently~\cite{DelDuca:2006hk} and a
de-correlation at the 25\% level was found for rapidity separations
$\Delta \eta_{jj}>4.2$, which would imply a corresponding reduction of $A$ 
in Eq.~(\ref{eqn:fit}). An even smaller effect was found in the NLO
calculation of Campbell et al.~\cite{Campbell:2006xx}, raising the
possibility that part of the de-correlation found in
Ref.~\cite{DelDuca:2006hk} is due to tagging jets originating from the
parton shower, which is generated flat in azimuthal angle. Considering
these effects we expect a reduction of 25\% or less in analyzing power
due to higher order effects. A quantitative determination of the
de-correlation requires further study, however.

\section{Conclusions}
\label{sec:conclusions}

Both sources of Higgs plus dijet events at the LHC, vector boson fusion
and gluon fusion, can provide important information on Higgs boson
properties. Vector boson fusion is sensitive to the couplings of the
Higgs boson to weak bosons, gluon fusion measures the effective $Hgg$
coupling, which, within the SM, is mostly induced by the $Htt$ Yukawa 
coupling. For a Higgs boson mass of 160~GeV and SM couplings, the decay
channel $H\to WW\to l^+ l^-\slashiv{p}_T$ provides a highly significant 
signal for both the vector boson fusion~\cite{VBF:H} and the gluon
fusion signal above backgrounds, which are dominated by top quark pair
production. The distinction of the two $Hjj$ channels is most easily
achieved, at the statistical level, by using characteristically
different distributions of the two tagging jets, mainly their rapidity
separation and the dijet invariant mass~\cite{Berger:2004pc}. A central
jet veto can be used to further enhance vector boson fusion over gluon
fusion events~\cite{VBF:H}. 

A further analysis of the structure of the $Htt$ Yukawa coupling, in
particular the question whether this coupling is CP-even or CP-odd or a
mixture of the two, is possible via the azimuthal angle correlation of
the two tagging jets. Taking into account the rapidity correlations of
the tagging 
jets with the beam directions, the sign of this azimuthal angle can be
defined~\cite{Hankele:2006ma} and the resulting full distribution in
azimuthal angle separation $\Delta\Phi_{jj}$ exhibits the relative
strength of CP-even and CP-odd couplings via a phase shift. The
$\Delta\Phi_{jj}$ distribution is sensitive to CP violation in the
Higgs sector. For the case analyzed here, $m_H=160$~GeV with SM size
production cross section in gluon fusion and SM-like decay branching
fractions, a highly significant 
measurement is expected with an integrated luminosity well below
100~fb$^{-1}$. A precise determination of the analyzing power requires a
full detector simulation, however, which is beyond the scope of the
present paper.

The methods presented here for the particular case of $m_H=160$~GeV, can
be extended to other Higgs boson masses. For $150\;{\rm GeV}\lsim
m_H\lsim 2m_Z$ the present analysis should be readily applicable, with
minor modifications, like in the transverse mass cut of
Eq.~(\ref{eqn:mtww}). Lower 
Higgs boson masses may require somewhat softer lepton $p_T$ cuts and
will eventually run into statistical problems due to the smaller
branching ratio for $H\to WW$ decay as compared to $m_H=160$~GeV. The
azimuthal angle correlations of the tagging jets are independent of the
specific Higgs decay mode, of course. This raises the
question whether $H\to\tau\tau$ or $H\to \gamma\gamma$ signals from
gluon fusion induced $Hjj$ events are observable at the LHC. Given the
much more difficult task of QCD background reduction as compared to
vector boson fusion studies which we have found in this paper for the ``easy''
$m_H=160$~GeV case, we expect such novel Higgs signals from gluon fusion
induced $Hjj$ events and the study of their azimuthal angle correlations
to be quite challenging, but worth considering.

\section{Acknowledgments}
We are grateful to Barbara J\"ager, Vera Hankele, Manuel B\"ahr and 
Michael Kubocz for many useful discussions.
This research was supported in part by the Deutsche Forschungsgemeinschaft
under SFB TR-9 ``Computational Particle Physics''. G.~K. greatfully 
acknowledges DFG support through the 
Graduiertenkolleg `` High Energy Physics and Particle Astrophysics''.


\end{document}